\documentclass[preprintnumbers,amsmath,amssymb,floatfix,10pt,prd,onecolumn,nofootinbib]{revtex4}
\usepackage{latexsym}
\usepackage{epsfig}
\usepackage{epstopdf}
\usepackage{graphicx,float}
\usepackage{amssymb}
\usepackage{amsmath}
\usepackage{dcolumn}
\usepackage{bm}
\usepackage{color}
\usepackage{comment}

\begin{document}
\title{\bf {Energy Constraints and Phenomenon of Cosmic Evolution in $f(T,B)$ Framework}}

\author{M. Zubair}
\email{mzubairkk@gmail.com;drmzubair@ciitlahore.edu.pk}\affiliation{Department
of Mathematics, COMSATS University Islamabad, Lahore Campus,
Pakistan}

\author{Saira Waheed}
\email{sairawaheed50@gmail.com}\affiliation{Prince Mohammad Bin Fahd University,
Al Khobar, 31952 Kingdom of Saudi Arabia.}

\author{M. Atif Fayyaz}
\email{atif_fayyaz@yahoo.com}\affiliation{Govt. Post Graduate College Kharian, Pakistan}

\author{Iftikhar Ahmad}
\email{iffi6301@gmail.com}\affiliation{Department
of Mathematics, COMSATS University Islamabad, Lahore Campus,
Pakistan}

\begin{abstract}

We investigate the cosmological evolution in a new modified teleparallel
theory, called $f(T,B)$ gravity, which is formulated by connecting both $f(T)$ and $f(R)$ theories with a boundary term $B$. Here, $T$ is the torsion scalar in teleparallel gravity and $R$ is the scalar curvature.
For this purpose, we assume flat Friedmann-Robertson-Walker (FRW)
geometry filled with perfect fluid matter contents.
We study two cases in this gravity: One is for a general function
of $f(T,B)$, and the other is for a particular form of it given by
the term of $-T+F(B)$. We also formulate the general energy constraints
for these cases.
Furthermore, we explore the validity of the bounds on the energy conditions by specifying different forms of$f(T,B)$ and $F(B)$ function obtained by the reconstruction scheme for de Sitter, power-law, the $\Lambda$CDM and Phantom cosmological models.
Moreover, the possible constraints on the free model parameters are examined with the help of region graphs. In addition, we explore the evolution of the effective equation of state (EoS) $\omega_{eff}$ for the universe and compare theoretical results with the observational data. It is found that the effective EoS represents the phantom phase or the quintessence one in the accelerating universe in all of the cases consistent with the observational data.\\


\end{abstract}

\maketitle

\date{\today}

\section{Introduction}

In the recent past, the development of a modified gravitational
framework that can successfully demonstrates the total cosmic matter
contents along with the complete history of cosmic evolution (from
big bang phenomenon till its final fate) is regarded as one of big
challenges. Although general theory of relativity (GR) is regarded
as a very successful theory that is consistent with the
observational results however it has some limitations on dark matter
and DE. So, the work on the modification of GR started just after
it's formulation. In this respect, Weyl made an effort for combining
the gravitation and electromagnetism soon after the final
presentation of GR \cite{1}. His effort was not as much successful
but it initiated the concept of gauge transformation and gauge
invariance and thus provided a basis for gauge theory \cite{2,3}.
Ten years later, Einstein \cite{4} made an effort by constructing
the structure of teleparallelism. In his effort, he used the concept
of tetrad, an orthogonal field based on the four-dimensional
space-time tangent space. As the tetrad has sixteen components,
therefore Einstein called this structure as a unification of
electromagnetism and gravity by relating six additional degrees of
freedom to electromagnetic field. Later on, it was also not proved
as problem free but the main idea of this theory is considered
important till date.

In this respect, Kaluza \cite{5} and Klein \cite{6} also proposed an
unified platform for gravity and electromagnetism, called
Kaluza-Klein theory. Further Cartan presented a successful
modification of GR, namely Einstein-Cartan theory, in which
spacetime involves both curvature and torsion \cite{7,8}. In this
theory, energy and momentum were the source of curvature, while the
spin was the source of torsion. In 1960, Moller \cite{9,10} made an
effort based on Einstein's idea to find a tensorial complex which is
invariant under coordinate transformation but not under local
Lorentz transformation. On the basis of Moller's work, Pellegrini
and Plebanski \cite{11} found a formulation of Lagrangian for
teleparallel gravity. In 1976, Cho \cite{12} found teleparallel
Lagrangian with the coefficient of anholonomy by replacing torsion
and it was invariant under local Lorentz transformation. In 1979,
Hayashi and Shirafuji \cite{13} made an effort to unify the concept
of teleparallelism with his earlier proposed gauge theory
\cite{14,15,16}.

Teleparallel gravity (TG) is a modified theory that is equivalent to
GR based on a different concept, like GR determines trajectories by
geodesics but TG consider torsion as a responsible candidate for
gravitation. Weitznb$\ddot{o}$ck suggested that it is possible to
choose such a connection for which the curvature vanishes and this
is regarded as a main idea behind teleparallel theory.
Interestingly, the field equations obtained through this formulation
are equivalent to that of GR \cite{17,18}. The interest in TG is
raising day by day over past few decades and many other extensions
of this theory has been proposed in literature \cite{19}. One of its
interesting modification is $f(T,T_G)$ gravity involving torsion
invariant T and contribution from a term $T_G$, the teleparallel
equivalent of the Gauss-Bonnet term \cite{20}. Another modification
involves general $f(T)$ function and its non-minimal coupling with
matter \cite{21}. Much work has been done on these teleparallel
frameworks and researchers obtained viable results for various
cosmological issues (for reviews on modified gravity and dark energy problem
to account for the late-time cosmic acceleration, see, e.g.,~\cite{Nojiri:2006ri, Nojiri:2010wj, Book-Capozziello-Faraoni, Capozziello:2011et, Bamba:2012cp, Joyce:2014kja, Koyama:2015vza, Bamba:2015uma, Cai:2015emx, Nojiri:2017ncd}).

$R=-T+B$ is the one of the basic equation of GR and its teleparallel
equivalent, where $R$ is the Ricci scalar, $T$ is the torsion scalar
and $B$ is a total derivative term which only depends on torsion,
named as boundary term. Thus Einstein-Hilbert action can be
represented by either using Ricci scalar or the torsion scalar as it
gives identical equations of motion. It is worthwhile to mention
here that variation of $f(R)$ Lagrangian (modification of GR)
results in fourth order field equations \cite{22,23} while $f(T)$
Lagrangian leads to second order field equations \cite{24}. As the
terms $T$ and $B$ are not invariant and hence under local Lorentz
transformation, this theory is no longer invariant \cite{25,26}, but
this issue can be resolved by taking the particular combination
$-T+B$ \cite{27}. So the $f(R)$ and $f(T)$ theories are not
equivalent but a relationship can be developed by using theories
based on $f(T,B)$ functions. In \cite{28}, Bahamonde et al. reconstructed some well known
models in the background of $f(T,B)$ gravity, discussing the thermodynamic properties and stability of such models.

The investigation of possible bounds on free parameters arising from
different DE models by making them consistent with the energy
conditions has always been a center of interest for the researchers.
Such constraints have already been explored in various gravitational
frameworks like $f(R)$ gravity, $f(T)$ theory, $f(G)$ theory, $f(R)$
gravity involving non-minimal interaction with matter,
$f(R,\mathcal{L}_m)$ gravity and Brans-Dicke theory \cite{29}. In
this context, Sharif and Saira \cite{30} considered FRW geometry
with perfect fluid matter contents in the most general scalar-tensor
gravity involving second-order derivatives of scalar field in field
equations and discussed the possible validity of energy conditions.
Sharif and Zubair \cite{31} have investigated the consistency of
these bounds in $f(R,T)$ gravity that involves Ricci scalar and
energy-momentum tensor trace. By using power law cosmology, they
also developed the stability criteria for this configuration.
Further, in another study, they examined some models of
$f(R,T,R_{\mu\nu}T^{\mu\nu})$ gravity using energy inequalities
\cite{32}. Zubair and Waheed studied the validity of energy bounds
for power law FRW cosmology in a modified theory involving
non-minimal coupling of torsion scalar and perfect fluid matter
\cite{21}. In another paper \cite{33}, the same authors explored the
compatibility of energy constraints in $F(T,T_G)$ gravity using two
different proposed models of $F(T,T_G)$ for FRW geometry with
perfect fluid matter and obtained interesting results.

Being motivated from the literature, in the present manuscript, we
investigate the possible constraints on free parameters using energy
condition approach in $f(T,B)$ gravity. In the next section, we
present a brief formulation of this theory and its resulting field
equations for perfect fluid FRW geometry. In section \textbf{III},
we formulate the energy constraints for a general $f(T,B)$ function
as well as its specific form $-T+F(B)$. Section \textbf{IV} is
devoted to study these energy bounds for four different $f(T,B)$
models obtained by reconstruction scheme namely: de sitter universe,
power law cosmology, $\Lambda$CDM model and phantom cosmology. Here
we discuss the validity of energy conditions and evolution of
effective equation of state (EoS) parameter using graphs. In the
next section, we discuss these energy bounds for particular forms of
$-T+F(B)$ constructed by reconstruction scheme using all four cases.
Last section concludes the whole discussion and highlights the major
results.

\section{Basic Formulation of $f(T,B)$ theory of gravity and FRW geometry}

In this section, we discuss the basic notions of $f(T,B)$ gravity
and construct the corresponding field equations for flat FRW
geometry with perfect fluid matter source. For this purpose, we
consider the action of a modified version of teleparallel theory
proposed in a recent study \cite{27} given as follows
\begin{align}
\label{fTB} S_{f(T,B)} =\frac{1}{\kappa^2} \int dx^{4}\,e\,f(T,B) +
S_{\rm m}\,,
\end{align}
where $f(T,B)$ is a general function of torsion scalar $T$ and
boundary term $B$. Here $e$ represents the determinant of tetrad and
$S_m$ denotes the action of ordinary matter. Actually, Ricci scalar
of Levi-civita in terms of torsion can be written in the form:
\begin{equation}
R(e)=-T+\frac{2}{e}\partial_{\mu}(eT^{\mu}),
\end{equation}
where the term in addition involving torsion vector $T^{\mu}$ is
considered as the boundary term given by
$B=\frac{2}{e}\partial_{\mu}(eT^{\mu})$. It has been proved that by
choosing $f=f(T)$ and $f=f(-T+B)=f(R)$, it is possible to recover
both $f(T)$ and $f(R)$ theories, respectively. The variation of the
above action with respect to tetrad results in the following set of
field equations
\begin{eqnarray}\label{0002}
2e\delta ^{\lambda}_{\nu} \Box f_{B}-2e \nabla ^{\lambda} \nabla
_{\nu}f_{B} +eBf_{B}\delta ^{\lambda}_{\nu}+4e \big[(\partial
_{\mu}f_{B}) +(\partial _{\mu}f_{T}) \big] S^{\mu
\lambda}_{\nu}+4e^{a}_{\nu}\partial _{\mu}(eS^{\mu \lambda}
_{a})f_{T} -4ef_{T}T^{\sigma}_{\mu\nu}S_{\sigma}^{\lambda
\mu}-ef\delta _{\nu}^{\lambda}=16 \pi e \tau ^{\lambda}_{\nu},
\end{eqnarray}
where $\tau _{\nu}^{\lambda} = e_{\nu}^{a}\tau _{a}^{\lambda}$ is
standard energy-momentum tensor and $\Box=\nabla ^{\mu} \nabla
_{\mu}$.

The line element for flat FRW geometry is given by
\begin{eqnarray}\label{0001}
ds^{2} = -dt^{2} + a(t)^{2}(dx^{2} + dy^{2} + dz^{2}) ,
\end{eqnarray}
where $a(t)$ is the expansion radius of universe. In these
coordinates, the tetrad field can be expressed as follows
\begin{equation}
e_{\mu }^{a}=\text{\textrm{diag}}\left( 1,a(t),a(t),a(t)\right)\,.
\label{fr.03}
\end{equation}%
We assume the source of ordinary matter as perfect fluid given by
\begin{eqnarray}\label{02}
T_{\mu\nu}=(\rho_m +p_m)u_{\mu}u_{\nu}-p_m g_{\mu\nu},
\end{eqnarray}
where $\rho_m$ and $p_m$ are ordinary matter density and pressure,
respectively.

Using (\ref{0001}) along with (\ref{02}), the corresponding field
equations (\ref{0002}) takes the following form
\begin{eqnarray}
&&-3H^{2}(3f_{B}+2f_{T})+3H\dot{f}_{B}-3\dot{H}f_{B}+\frac{1}{2}f(T,B)=\kappa
^{2}\rho _{m},\\&&
-3H^{2}(3f_{B}+2f_{T})-\dot{H}(3f_{B}+2f_{T})-2H\dot{f}_{T}+\ddot{f}_{B}+\frac{1}{2}f(T,B)=-\kappa
^{2}p _{m}.
\end{eqnarray}
Here $H=\frac{\dot{a}}{a}$ is the Hubble parameter and dot
represents derivative with respect to $t$. In terms of effective
energy and pressure terms, the above set of equations can be
rewritten as
\begin{eqnarray}
&& 3H^{2} = \kappa ^{2}_{eff}(\rho _{eff}), \\ && 2\dot{H} = -\kappa
^{2}_{eff}(\rho _{eff}+p_{eff}),
\end{eqnarray}
where
\begin{eqnarray}
&&\rho _{eff} =\rho _{m}+ \frac{1}{\kappa ^{2}}
\bigg[-3H\dot{f}_{B}+(3\dot{H}+9H^{2})f_{B}-\frac{1}{2}f(T,B)\bigg],\\&&
p_{eff}=p_{m}+ \frac{1}{\kappa ^{2}}
\bigg[\frac{1}{2}f(T,B)+\dot{H}(2f_{T}-3f_{B})-2H\dot{f}_{T}-9H^{2}f_{B}+\ddot{f}_{B}\bigg].
\end{eqnarray}

\section{Energy Conditions}

In this section, firstly, we present a general discussion on energy
condition bounds in GR and then formulate specifically these
constraints for the present case. The origin of these conditions
emerges from the Raychaudhuri equation along with the condition that
the gravity is attractive. For this, consider the tangent vector
field $u^{\mu}$ that is congruent to timelike geodesics in a
spacetime manifold endowed with a metric $g_{\mu \nu}$. Then
Raychaudhuri's equation is given by
\begin{eqnarray}
\frac{d\theta}{d\tau}=-\frac{1}{3}\theta^{2}- \sigma_{\mu \nu}
\sigma^{\mu \nu} - \omega_{\mu \nu}\omega_{\mu \nu}-R_{\mu
\nu}u^{\mu}u^{\nu},
\end{eqnarray}
where terms $R_{\mu \nu},~\theta$ and $\sigma ^{\mu \nu}$ denote the
Ricci tensor, expansion, and shear scalars, respectively. Also,
$\omega_{\mu\nu}$ is the rotation associated to the congruence
defined by the vector field $u^{\mu}$. The above equation deals with
the geometry and has no reference to gravitational field equations.
Since the GR field equations relate Ricci tensor $R_{\mu\nu}$ to the
energy-momentum tensor $T_{\mu\nu}$, therefore the combination of
Einstein's and Raychaudhuri's equations can be used to restrict the
energy-momentum tensor on physical grounds. As the shear is spatial
tensor and $\sigma^{2}\equiv\sigma_{\mu\nu}\sigma^{\mu\nu} \geq 0$,
so from Raychaudhuri's equation, the condition for attractive
gravity $\frac{d\theta}{d\tau}<0$ reduces to
$R_{\mu\nu}u^{\mu}u^{\nu}\geq0$ for any hypersurface orthogonal
congruences ($\omega\equiv 0$). Consequently, by using Einstein's
equation, it can be written as
\begin{eqnarray} \label{01}
R_{\mu\nu}u^{\mu}u^{\nu}=\left(T_{\mu\nu}-\frac{T}{2}g_{\mu\nu}\right)u^{\mu}u^{\nu}\geq0.
\end{eqnarray}
Here we consider $8\pi G=c=1$. Further, symbols $T_{\mu\nu}$ and $T$
represent the energy-momentum tensor and its trace, respectively.

Equation (\ref{01}) gives the strong energy condition (SEC) in a
coordinate-invariant way in terms of $T_{\mu\nu}$. So SEC in the
context of GR shows that the gravity is attractive. For a perfect
fluid matter source (\ref{02}), Eq.(\ref{01}) takes the form as
$\rho_m+3p_m\geq 0$. Other energy constraints namely null energy
condition (NEC), weak energy condition (WEC) and dominant energy
conditions (DEC) can be written as
$$NEC: \rho_m+p_m \geq,~ WEC: \rho_m\geq 0,~ \rho_m+p_m\geq 0,~ DEC: \rho_m\geq 0,~ \rho_m\pm p_m \geq
0.$$

The derivation of energy conditions for $f(T,B)$ gravity can be done
in a similar pattern. In the present work, we consider that the
total matter contents acts as a perfect fluid. So these conditions
can be obtained by just replacing $\rho_m$ with $\rho _{eff}$ and
$p_m$ with $p_{eff}$ as follows
\begin{eqnarray} \label{001}
NEC&:&~\rho_{eff}+p_{eff}=\rho_{m}+p_{m}+\frac{1}{\kappa^{2}}\bigg[-3H
\dot{f}_{B}+3\dot{H}f_{B}+\dot{H}(3f_{B}+2f_{T})-2H\dot{f}_{T}+\ddot{f}_{B}\bigg]\geq
0,\\\label{002} WEC&:&~\rho_{eff}=\rho_{m}+\frac{1}{\kappa
^{2}}\bigg[-3H\dot{f}_{B}+3\dot{H}f_{B}-\frac{1}{2}f(T,B)\bigg]\geq
0,~\rho_{eff}+p_{eff}\geq0,\\\nonumber
SEC&:&~\rho_{eff}+3p_{eff}=\rho_{m}+3p_{m}+\frac{1}{\kappa^{2}}\bigg[-3H\dot{f}_{B}+12\dot{H}f_{B}
+f(T,B)+6\dot{H}f_{T}-6H\dot{f}_{T}+3\ddot{f}_{B}\bigg]\geq
0,\\\label{003}&&\rho_{eff}+p_{eff}\geq0,\\\nonumber
DEC&:&~\rho_{eff}-p_{eff}=\rho_{m}-p_{m}+\frac{1}{\kappa^{2}}\bigg[-3H\dot{f}_{B}-2\dot{H}f_{T}
+2H\dot{f}_{T}-\ddot{f}_{B}-f(T,B)\bigg]\geq 0,\\\label{004}
&&\rho_{eff}\geq0,~\rho_{eff}+p_{eff}\geq0.
\end{eqnarray}
Inequalities (\ref{001})-(\ref{004}) represents the null, weak,
strong and dominant energy conditions in the context of $f(T,B)$
gravity for $FRW$ spacetime.

For flat FRW spacetime, the torsion scalar and the boundary term
takes the form:
\begin{eqnarray}\nonumber
T=6H^2, \quad B=6(\dot{H}+3H^2).
\end{eqnarray}
Also, for FRW line element, deceleration, jerk and snap cosmological
parameters are given by
\begin{equation}\nonumber
q=-\frac{1}{H^{2}}\frac{\ddot{a}}{a},\quad
j=\frac{1}{H^{3}}\frac{\dddot{a}}{a}, \quad
s=\frac{1}{H^{4}}\frac{\ddddot{a}}{a}
\end{equation}
The expressions of $T,~B$ and $H$ along with their derivatives in
terms of these cosmological parameters can be written as
\begin{eqnarray}\nonumber
T&=&6H^{2},\quad \dot{T}=-12H^{3}(1+q), \quad B=6H^{2}(2 - q),\quad
\dot{B}=6H^{3}(j-3q-4),\\\nonumber H&=&\frac{\dot{a}}{a}, \quad
\dot{H}=-H^{2}(1 + q),\quad \ddot{H}=(j+3q+2)H^{3},\quad
\dddot{H}=H^{4}(s-4j-12q-3q^{2}-6).
\end{eqnarray}
Using the above definitions, the energy conditions
(\ref{001})-(\ref{004}) will take the form as follows
\begin{eqnarray}\nonumber
NEC&:&\kappa^{2}(\rho_{m}+p_{m})-2H^{2}(1+q)(3f_{B}+f_{T})-18H^{4}(j-3q-4)f_{BB}+
36H^{4}(1+q)\\\nonumber &-& 12H^{4}(j-3q-4)f_{BT}-
24H^{4}(1+q)f_{TT}+36H^{6}(j-3q-4)^{2}f_{BBB}\\\label{005}&-&144H^{6}(1+q)(j-3q-4)f_{BBT}+144H^{6}(1+q)^{2}f_{BTT}\geq
0,\\\nonumber
WEC&:&\kappa^{2}\rho_{m}-18H^{4}(j-3q-4)f_{BB}+36H^{4}(1+q)f_{BT}-3H^{2}(1+q)f_{B}-\frac{1}{2}f(T,B)\geq
0,\\\label{006}&&\rho_{eff}+p_{eff}\geq0,\\\nonumber
SEC&:&\kappa^{2}(\rho_{m}+3p_{m})-18H^{4}(j-3q-4)f_{BB}+36H^{4}(1+q)f_{BT}-12H^{2}(1+q)f_{B}-6H^{2}(1+q)f_{T}\\\nonumber
&-&36H^{4}(j-3q-4)f_{BT}+72H^{4}(1+q)f_{TT}+108H^{6}(j-3q-4)^{2}f_{BBB}-72H^{6}(1+q)(j-3q-4)f_{BBT}\\\label{007}
&+&432H^{6}(1+q)f_{BTT}+f(T,B)\geq 0, \quad
\rho_{eff}+p_{eff}\geq0,\\\nonumber
DEC&:&\kappa^{2}(\rho_{m}-p_{m})-18H^{4}(j-3q-4)f_{BB}
+12H^{4}(1+q)f_{BT} +2H^{2}(1+q)f_{T} -24H^{4}(1+q)f_{TT}\\\nonumber
&&- 36H^{6}(j-3q-4)f_{BBB}-144H^{6}(1+q)(j-3q-4)f_{BBT}
+144H^{6}(1+q)^{2}f_{BTT}-f(T,B)\geq
0,\\\label{008}&&\rho_{eff}+p_{eff}\geq0,\quad \rho _{eff}\geq 0.
\end{eqnarray}

\section{Evolution of Energy condition Bounds using Reconstructed Models of $f(T,B)$}

In this section, we explore the evolution of energy condition bounds
(\ref{005})-(\ref{008}) using some interesting $f(T,B)$ models
obtained by reconstruction scheme. For this purpose, we consider
four well-known cosmological models namely de Sitter (dS), power
law, $\Lambda$CDM and phantom models of cosmos.

\subsection{de Sitter universe model}

The dS solutions are considered as one of the most fascinating
models in cosmology for explaining the current accelerated cosmic
epoch. The dS model is described by the exponential scale factor
defined as $a(t)=a_{0}e^{H_{0}t}$, where $a_0$ and $H_0$ are the
present values of scale factor and Hubble parameter, respectively.
Further, the torsion scalar and the boundary term takes the form:
$$T= 6H_{0}^{2},~ B= 18H_{0}^{2}.$$ Also, we
consider the perfect fluid matter source satisfies constant EoS
parameter given by $\omega_m=\frac{p_m}{\rho_m}$. Thus,
\begin{eqnarray}\nonumber
\rho=\rho_{0}e^{-3(1+\omega_m)H_{0}t},\quad
\rho_{m}=\rho_{0}a(t)^{-3(\omega_m+1)},\quad p_{m}=\omega_m\rho_{m}.
\end{eqnarray}
Consequently, the reconstruction technique leads to the following
form of $f(T,B)$ model given by
\begin{eqnarray}\label{009}
f(T,B)=2(\kappa^{2}\rho_{0}+2K)+f_{0}e^{\frac{B}{18H_{0}^{2}}}+\widetilde{f}_{0}e^{\frac{T}{12H_{0}^{2}}}.
\end{eqnarray}
Here $f_{0}$ and $\widetilde{f}_{0}$ are integration constants,
while $K$ is the constant of separation. Introducing Eq.(\ref{009})
in the set of energy conditions (\ref{005}) and (\ref{008}), we get
\begin{eqnarray}\nonumber
NEC&:& e(-2+j^{2}-3q+9q^{2}-j(17+6q))f_{0}+54(1 +
q)(108H_0^{4}-\sqrt{e}\widetilde{f}_{0})\geq 0,\\\nonumber WEC&:&
e(8 + j)f_{0}+9\left(4K+\sqrt{e}\tilde{f}_{0}\right)\geq 0,\quad NEC
\geq 0.
\end{eqnarray}
Here, each of these inequalities depend on the values of
$j,~q,~H_0,~f_0,~K$ and $\tilde{f_0}$. For graphical illustration,
we consider the recent values of cosmic parameters $H_0$, $q$ and
$j$. proposed by Capozziello et al. \cite{34}. These values are
$H_0=0.718,~q_0=-0.64$ and $j_0=1.02$. Also, we assume that the
energy constraints hold for ordinary matter source and
$\kappa^2=8\pi$, gravitational coupling constant, is a positive
quantity, therefore we only discuss the inequalities for DE source.
Here we explore the possible ranges of free parameters $f_0$ and
$\tilde{f_0}$ for which the energy constraints are satisfied. The
graphical behavior of these inequalities is given in left plot of
Figure \textbf{1}.
\begin{figure}
\center\epsfig{file=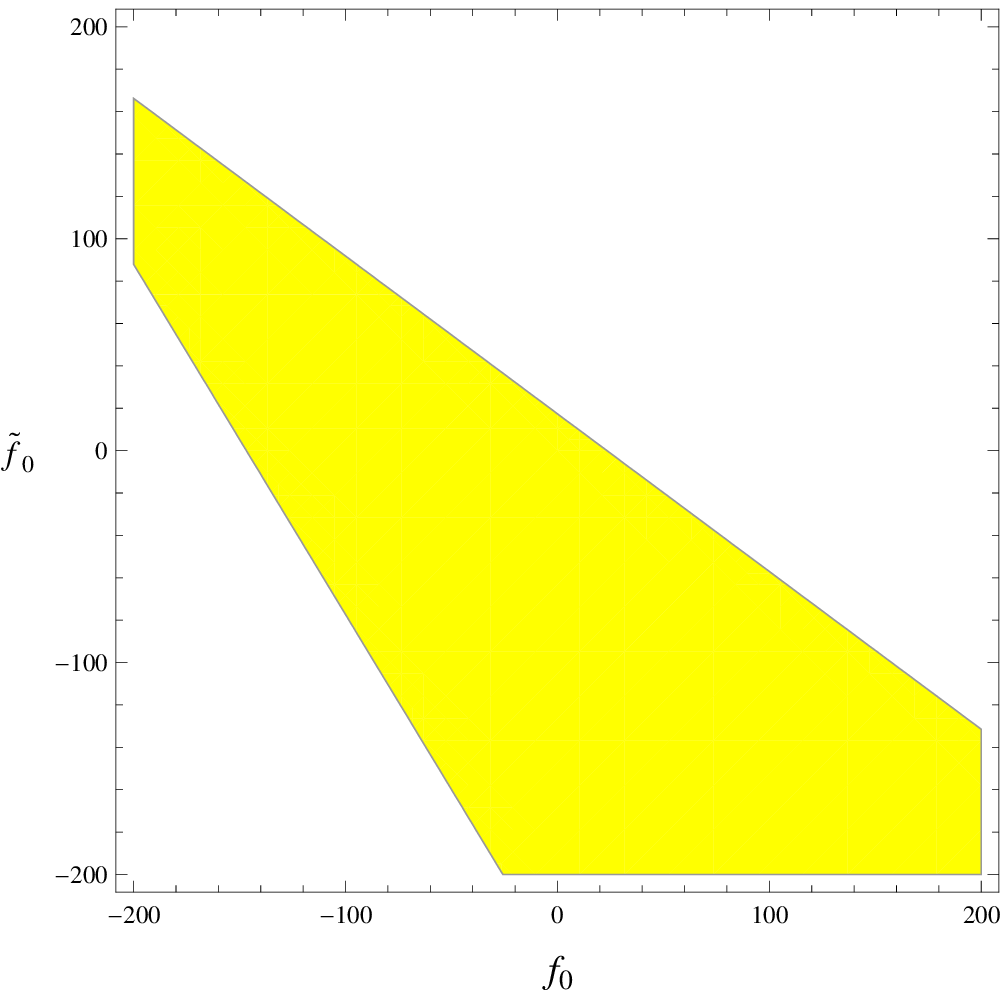, width=0.35\linewidth}
\epsfig{file=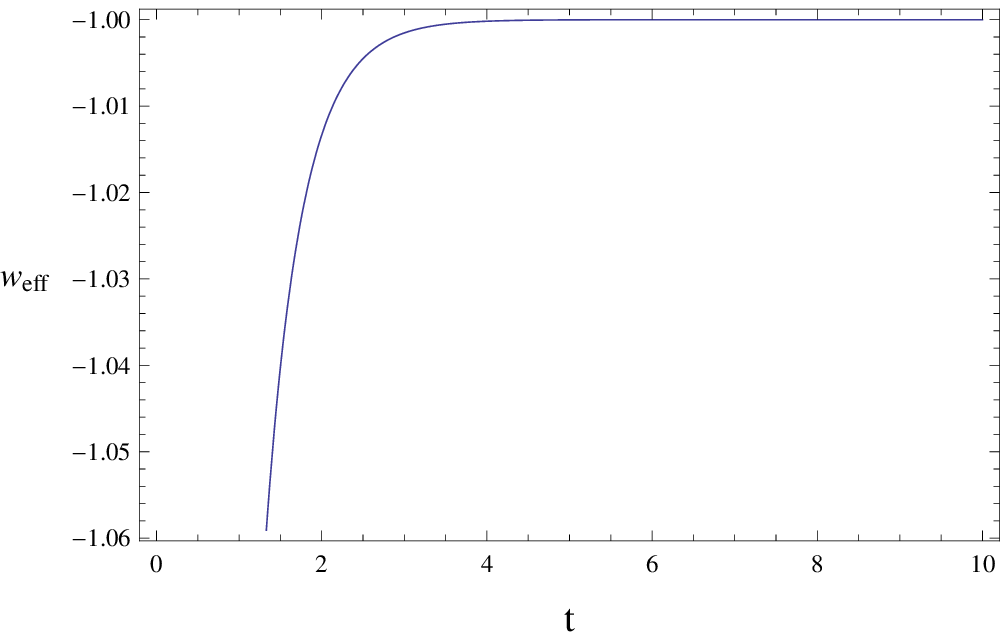, width=0.35\linewidth}\caption{Left region
plot shows validity region for WEC and NEC, while Right plot
indicates the evolution of effective EoS parameter versus cosmic
time. Here $K=1,~\omega_m=0,~\rho_0=1$. In right plot, in addition
to this, we consider $\tilde{f}_0=-2$.}
\end{figure}
It can be seen that both WEC and NEC remains valid when the free
parameters are constrained approximately within these limits:
$-25\leq f_0\leq 200,~ 88\leq\tilde{f_0}\leq 170$. These obtained
ranges of free parameters are consistent with the one obtained in
\cite{28}. In this case, the effective EoS parameter is given by
\begin{eqnarray}\nonumber
\omega_{eff}=\frac{\omega_m\rho_0
e^{-3(1+\omega_m)H_0t}+\rho_0+\frac{2K}{\kappa^2}+\frac{\tilde{f_0}e^{T/12H_0}}{2\kappa^2}}{\rho_0
e^{-3(1+\omega_m)H_0t}+\rho_0-\frac{2K}{\kappa^2}-\frac{\tilde{f_0}e^{T/12H_0}}{2\kappa^2}}.
\end{eqnarray}
Its graphical illustration of $\omega_{eff}$ is given in right plot of Figure
\textbf{1}. It is clear that the effective EoS approaches to $-1$ for late times of the universe.

\subsection{Power law solutions}

Here we will investigate the compatibility of WEC and NEC for
$f(T,B)$ model reconstructed using power law as expansion factor
\cite{28}. Further, we explore the behavior of effective EoS
parameter. Let us consider a model described by a power-law scale
factor given by
$$a(t)=\left(\frac{t}{t_{0}}\right)^{h},$$
where $t_{0}$ is some fiducial time and $h$ denotes a constant value
greater than zero. Such solutions help to explain different phases
of cosmic history by taking some specific values of $h$ like
matter-matter dominated epoch corresponds to $h=\frac{2}{3}$, while
radiation-dominated era relates to $h=\frac{1}{2}$. Also, $h>1$
predicts a late-time accelerating stage of universe. For the above
scale factor, the scalar torsion and boundary can be written as
follows
$$T=\frac{6h^{2}}{t^{2}},~ B=\frac{6h(3h-1)}{t^{2}}.$$
By assuming that the ordinary matter contents satisfies the EoS
parameter given by $\omega_m=\frac{p_m}{\rho_m}$, we get
$$\rho_m=\rho_{0}\left(\frac{t}{t_{0}}\right)^{-3h(1+\omega)},~
\rho_{m}=\rho_{0}(a(t))^{-3(\omega_m+1)},~ p_{m}=
\omega_m\rho_{m}.$$ Under all these assumptions, the possible
reconstructed form of $f(T,B)$ function describing power-law
cosmology is given by
\begin{eqnarray}\label{010}
f(T,B)=\sqrt{T}C_{1}+B^{\frac{1}{2}(1-3
h)}C_{2}+BC_{3}+\frac{2^{1-\frac{3}{2}h(1+\omega_m)}3^{-\frac{3}{2}h(1+\omega_m)}
\kappa^{2}(\frac{\sqrt{T}t_{0}}{h})^{3h(1+\omega_m)}\rho_{0}}{1-3h(1+\omega_m)}.
\end{eqnarray}
Introducing Eq.(\ref{010}) in the energy conditions (\ref{005}) and
(\ref{006}), we get
\begin{eqnarray}
NEC&:&\nonumber\frac{1}{6}H^{4}(1+q)(216+\frac{\sqrt{6}t^{3}C_{1}}{h^{3}}-\frac{2^{-\frac{5}{2}-\frac{3h}{2}}\times
3^{\frac{1}{2}-\frac{3h}{2}}(1+ 3h)H^{4}(-4+j-3q)\left(\frac{h(-1+3
h)}{t^{2}}\right)^{\frac{1}{2}(-1-3h)}t^{2}C_{2}}{h}\\\nonumber
&-&\frac{{2^{-\frac{7}{2}-\frac{3h}{2}}}\times3^{\frac{1}{2}-\frac{3h}{2}}
(1+h)(1+3h)H^{6}(-4+j-3q)^{2}\left(\frac{h(-1+3h)}{t^{2}}\right)^{\frac{1}{2}(-1-3h)}t^{6}C_{2}}{(1-3h)^{2}h^{3}}\\\label{*1}
&&-\frac{H^{2}(1+q)(\sqrt{6}tC_{1}+3(-6^{\frac{1}{2}-\frac{3h}{2}}
\left(\frac{h(-1+3h)}{t^{2}}\right)^{\frac{1}{2}(1-3h)}t^{2}C_{2}+12hC_{3}))}{6h}\geq 0,
\\\nonumber
WEC&:&\sqrt{6}htC_{1}+
6^{\frac{1}{2}-\frac{3h}{2}}\left(\frac{h(-1+3h)}{t^{2}}\right)^{\frac{1}{2}(1-3h)}t^{2}C_{2}\\\nonumber
&+&\frac{2^{-\frac{3}{2}(1+h)}\times
3^{\frac{1}{2}-\frac{3h}{2}}(1+3h)H^{4}(-4 + j - 3
q)\left(\frac{h(-1 + 3 h)}{t^{2}}\right)^{\frac{1}{2}(-1 - 3
h)}t^{4}C_{2}}{h}\\\nonumber &+& 6h(-1 + 3 h)C_{3}+3H^{2}(1 +
q)t^{2}\left(-\frac{6^{-\frac{1}{2}-\frac{3h}{2}}\left(\frac{h(-1+3h)}{t^{2}}\right)^{\frac{1}{2}(1-3h)}t^{2}C_{2}}{h}
+2C_{3}\right)\geq 0,~\rho _{eff}+p_{eff}\geq0.\\\label{*2}
\end{eqnarray}
These conditions involve parameters like $q,~C_1,~C_2,~C_3,~h,~j$
and the cosmic time $t$. We explore the validity of these bounds by
fixing all other free parameters except $C_1$ and also by varying
cosmic time. The graphical behavior of these constraints is given in
the left panel of Figure \textbf{2}. Since $h>1$ indicates
accelerated expanding late cosmic stages therefore we have chosen
values greater than 1. It is seen that by taking greater values of
$h$ and $C_1$, the validity region of these energy bounds can be
extended with increasing cosmic time. Also, it is clear from the
graph that for very few negative $C_1$ values with small $t$ satisfy
these energy bounds.
\begin{figure}
\center\epsfig{file=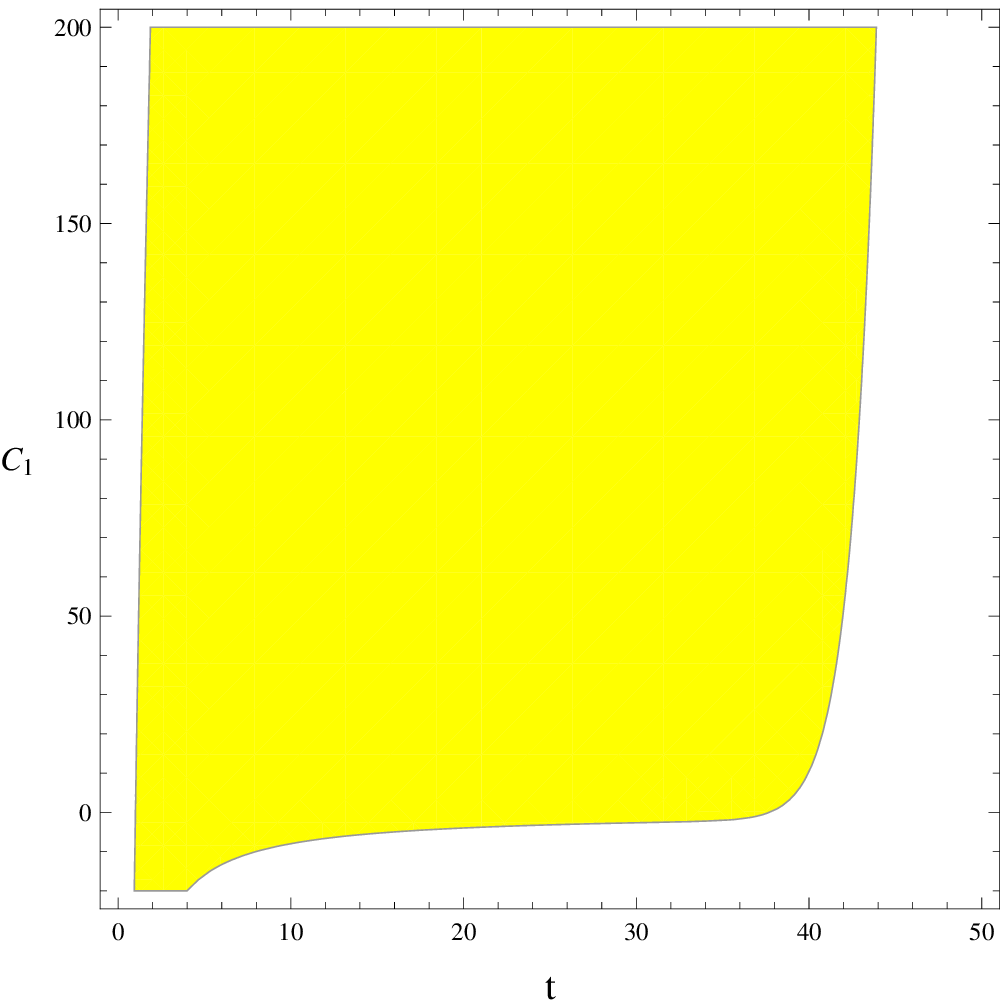, width=0.35\linewidth}
\epsfig{file=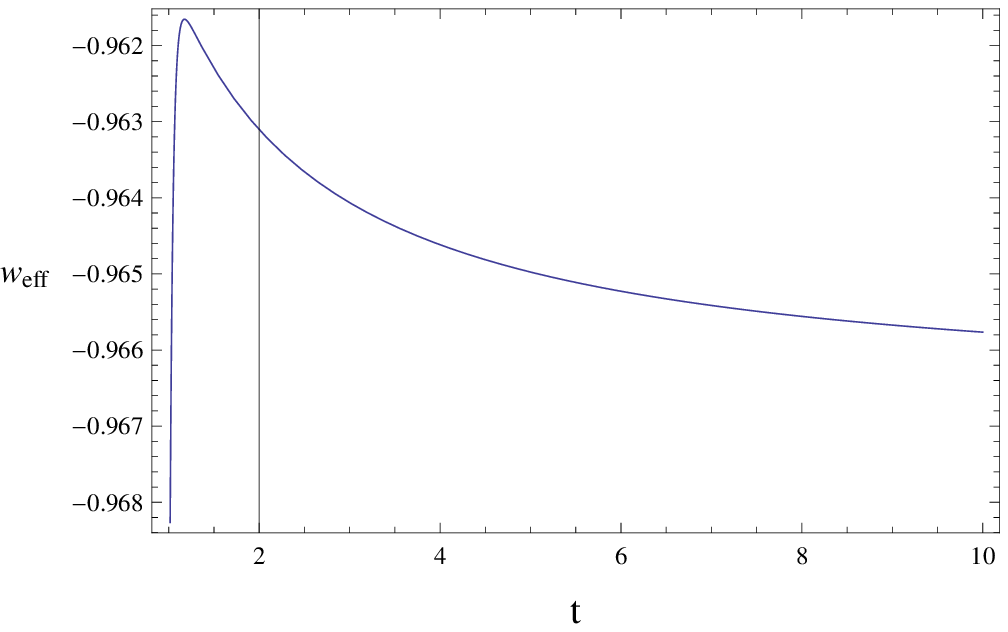, width=0.35\linewidth}\caption{Left panel
shows validity region for WEC and NEC while Right plot indicates the
evolution of effective EoS parameter versus cosmic time. Here
$\omega_m=0,~t_0=\rho_0=C_2=C_3=1$ and $h=10$. Also, in right plot,
we take $C_1=100$.}
\end{figure}

Further, the effective EoS parameter in this case is given by
\begin{eqnarray}
\omega_{eff}&=&\omega_m\rho_m+\frac{1}{8\pi}[\frac{1}{2}(\sqrt{T}C_{1}+B^{\frac{1}{2}(1-3
h)}C_{2}+BC_{3}+\frac{2^{1-\frac{3}{2}h(1+\omega_m)}3^{-\frac{3}{2}h(1+\omega_m)}8\pi
(\frac{\sqrt{T}t_{0}}{h})^{3h(1+\omega_m)}\rho_{0}}{1-3h(1+\omega_m)})\\\nonumber
&-&\frac{h}{t^2}(\frac{C_1}{\sqrt{T}}+\frac{2^{2-3/2h(1+\omega_m)}3^{3/2h(1+\omega_m)}}{1-3h(1+\omega_m)}
\left(\frac{\sqrt{T}t_0}{h}\right)^{3h(\omega_m+1)}\frac{t_0}{2h\sqrt{T}}-1.5C_2(1-3h)B^{1/2(1-3h)-1}-3C_3)
\\\nonumber&-&2\frac{h}{t}({\frac{C_1}{4}T^{-\frac{3}{2}}-\frac{2^{2-\frac{3}{2}h(1+\omega_m)}3^{\frac{3}{2}h(1+\omega_m)}8\pi\rho_0}{(1-3h(1+\omega_m))}
\frac{3h(1+\omega_m)}{2}(\frac{t_0}{h})^{3h(1+\omega_m)}(\frac{3}{2}h(1+\omega_m)-1)T^{\frac{3}{2}h(1+\omega_m)-2}}(\frac{12h^2}{t^3}))
\\\nonumber&-&9(\frac{h}{t})^2(\frac{1}{2}C_2(1-3h)B^{\frac{1}{2}(1-3h)-1}+C_3)+(\frac{1}{2}(1-3h)(\frac{1}{2}(1-3h)-1)(\frac{1}{2}(1-3h)-2)C_2
B^{\frac{1}{2}(1-3h)-3})\\\nonumber&\times&(-\frac{12h(3h-1)}{t^3})^2
+(\frac{1}{2}(1-3h)(\frac{1}{2}(1-3h)-1)C_2B^{\frac{1}{2}(1-3h)-2})(\frac{36h(3h-1)}{t^4})]
\\\nonumber&\times&[\rho_m+\frac{1}{8\pi}\{\frac{3h}{2t}(1-3h)(\frac{1}{2}(1-3h)-1)C_2B^{\frac{1}{2}(1-3h)-2)\frac{12h(3h-1)}{t^2}}
+(-\frac{3h}{t^2}+\frac{9h^2}{t^2})\\\nonumber&\times&(\frac{C_2}{2}(1-3h)B^{\frac{1}{2}(1-3h)-1)}+C_3)-\frac{1}{2}(\sqrt{T}C_{1}
+B^{\frac{1}{2}(1-3h)}C_{2}+BC_{3}\\\nonumber&+&\frac{2^{1-\frac{3}{2}h(1+\omega_m)}3^{-\frac{3}{2}h(1+\omega_m)}
8\pi(\frac{\sqrt{T}t_{0}}{h})^{3h(1+\omega_m)}\rho_{0}}{1-3h(1+\omega_m)}))\}]^{-1}.
\end{eqnarray}
The evolution of this parameter is given in the right panel of
Figure \textbf{2}. Clearly $-1\leq\omega_{eff}\leq0$ represents the
accelerated expanding stages of cosmic evolution.

\subsection{$\Lambda$CDM Cosmology}

In this section, we will use the reconstructed $f(T,B)$ function for
a $\Lambda CDM$ cosmological evolution without including any
cosmological constant term in this modified gravity. For this
purpose, we take $f(T,B)$ as obtained in \cite{28}
\begin{eqnarray}\label{011}
f(T,B)=BC_{1}+C_{1}\sqrt{T}+\frac{C_{2}}{\sqrt{B}}-\frac{2^{-\omega_m}\kappa
^{2}\rho_{0}}{2\omega_m+1}\left(\frac{T}{3a_{0}^{3}l}\right)^{\omega_m+1}.
\end{eqnarray}
Introducing Eq.(\ref{011}) in the energy conditions (\ref{005}) and
(\ref{006}), we get
\begin{eqnarray}\nonumber
NEC&:& 93312H^{5}(1 + q)-15552H^{3}(1 +
q)C_{1}+\sqrt{2}(64+22j-5j^{2}+6q+30jq-45q^{2})C_{2}\geq
0,\\\nonumber WEC&:&72 H^{2}\left(\sqrt{6}+6H(4 +
q)\right)C_{1}+\sqrt{2}\left(6 + j - 5 q\right)C_{2}\geq 0.
\end{eqnarray}
Here $a(t)=a_0e^{N},~H^2=g=le^{-3N},~l=\frac{8\pi}{3}$. Further, the
torsion scalar and boundary terms turn out to be $T=6le^{-3N}$ and
$B=9le^{-3N}$. It is worthwhile to mention here that $N=-\ln(1+z)$
is called e-folding parameter and $z$ is called redshift function.
It further yields the relationship: $a(t)=\frac{a_0}{1+z}$. The
graphical illustration of these energy bounds in terms of validity
region is given by left panel of Figure \textbf{3}. Clearly, these
constraints are satisfied for positive $C_1$ values only. It can
also be seen that there are few values $C_1>50$ with small values of
red shift function (large values of scale factor), where these
constraints are no longer valid.

Also, the effective EoS parameter is given by
\begin{eqnarray}\nonumber
\omega_{eff}&=&[\omega_m\rho_m+\frac{1}{8\pi}\{\frac{1}{2}(BC_{1}+C_{1}\sqrt{T}+\frac{C_{2}}{\sqrt{B}}-\frac{2^{-\omega}\kappa
^{2}\rho_{0}}{2\omega+1}\left(\frac{T}{3a_{0}^{3}l}\right)^{\omega+1})-\frac{3}{2}le^{-3N}(2(\frac{C_3}{2\sqrt{T}}\\\nonumber
&-&\frac{2^{-\omega}8\pi\rho_0}{2\omega_m+1}(\omega_m+1)\frac{1}{3a_0^3l}\left(\frac{T}{3a_0^3l}\right)^{\omega_m})
-3(C_1-\frac{C_2}{2B^{\frac{3}{2}}}))-2H((-\frac{C_3}{4T^{\frac{3}{2}}}\\\nonumber
&-&\frac{2^{-\omega}8\pi\rho_0\omega_m}{2\omega_m+1}(\omega_m+1)\frac{1}{(3a_0^3l)^2}
\left(\frac{T}{3a_0^3l}\right)^{\omega_m-1})H(-18le^{-3N}))
-9H^2(C_1-\frac{C_2}{2B^{\frac{3}{2}}})\\\nonumber
&-&\frac{243}{4}C_2l^{-1}He^{3N}\}][\rho_m+\frac{1}{8\pi}\{3H(\frac{81}{4}C_2lHB^{-\frac{5}{2}}e^{-3N})+(-3\frac{3}{2}le^{-3N}+9le^{-3N})\\\nonumber
&-&\frac{1}{2}(BC_{1}+C_{1}\sqrt{T}+\frac{C_{2}}{\sqrt{B}}-\frac{2^{-\omega}\kappa
^{2}\rho_{0}}{2\omega+1}\left(\frac{T}{3a_{0}^{3}l}\right)^{\omega+1})\}]^{-1}.
\end{eqnarray}
\begin{figure}
\center\epsfig{file=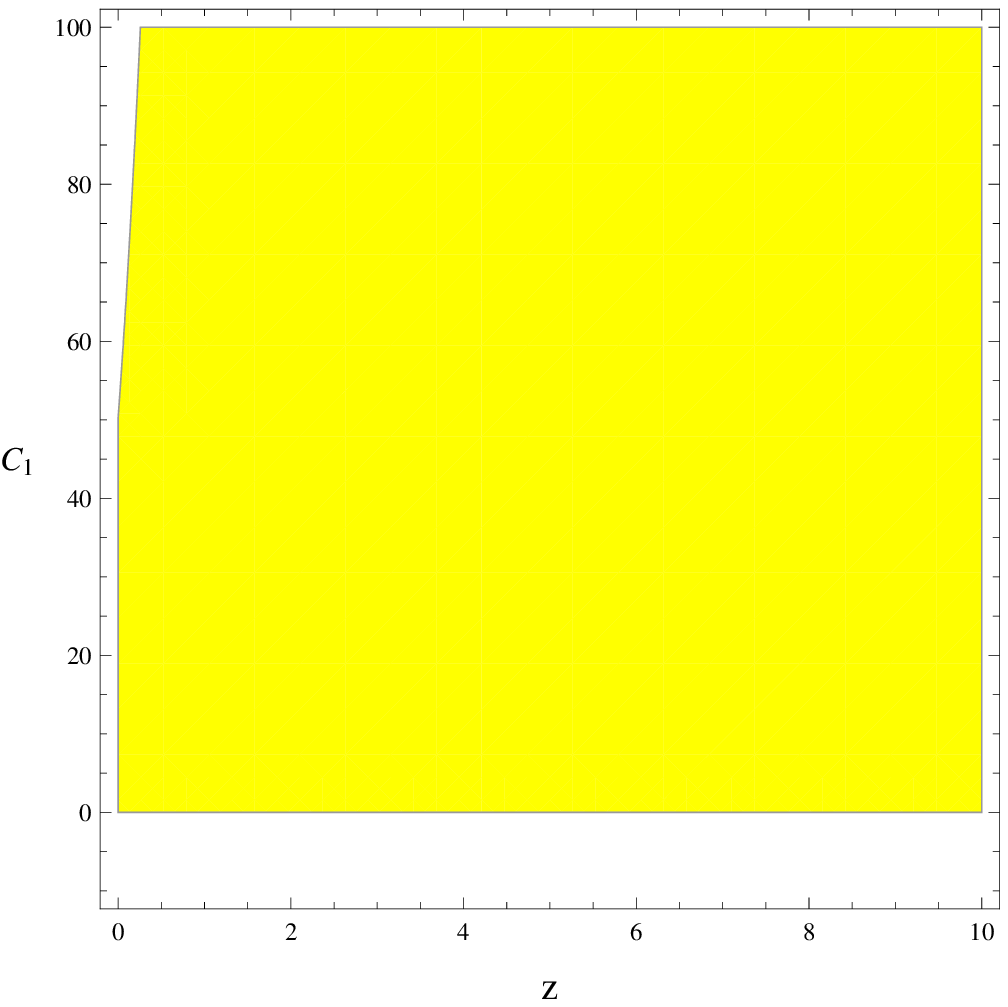, width=0.35\linewidth}
\epsfig{file=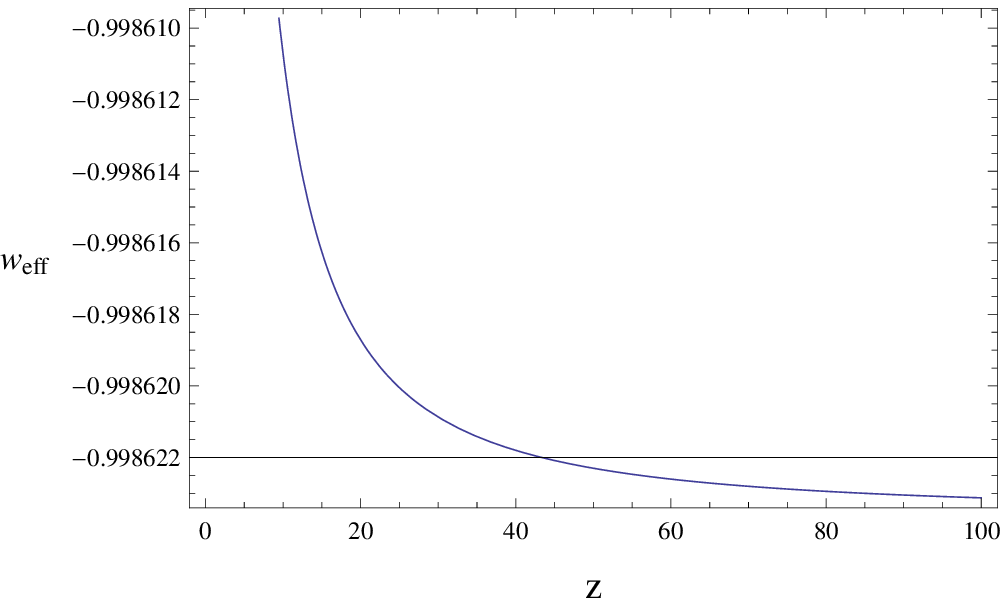, width=0.35\linewidth}\caption{Left region
plot shows validity region for WEC and NEC while Right plot
indicates the evolution of effective EoS parameter versus cosmic
time. For left graph, we consider $q=-.64,~j=1.02,~l=\frac{8\pi}{3}$
and $C_2=10$. For right plot, we take
$\omega_m=0,~a_0=1=\rho_0,~C_1=1000,~C_2=C_3=10$.}
\end{figure}
The dynamics of effective EoS parameter is provided in the right
penal of Figure \textbf{3}. It indicates that the universe model is
in quintessence stage of its evolution.

\subsection{Phantom Universe Model}

In this section, we will discuss the evolution of energy condition
bounds for reconstructed phantom model of universe \cite{28}. This
phantom universe model is given by
\begin{equation}\label{p1*}
f(T,B)=C_1B^{\frac{3+m}{2m}}+C_2B+16\pi b_0+C_3\sqrt{T}-\frac{8\pi
b_1T}{3h_0^2}-\frac{\sqrt{8\pi\frac{8}{3}}b_2(m+1)T^{\frac{5}{2}}}{15h_0^5},
\end{equation}
where $C_1,~C_2,~C_3,~b_0,~b_1,~m$ and $h_0$ are all constants. Here
the ordinary energy density and Hubble parameter are given by
$\rho_m=b_0+b_1e^{2mN}+\frac{96(m+1)}{5}b_2e^{5mN};~m\neq-3$ and
$H=\sqrt{g}=h_0e^{mN}$, respectively. Further $N=\ln(\frac{1}{1+z})$
and is called e-folding parameter and $z$ is the red shift
parameter. For this model (\ref{p1*}), the WEC (\ref{005}) and NEC
(\ref{006}) are given by
\begin{eqnarray}\nonumber
&&8\pi(1+\omega_m)\left(b_0+b_1e^{2mN}+\frac{96(m+1)}{5}b_2e^{5mN}\right)-h_0^2e^{2mN}(1+q)
\{3\left(C_2+C_1(\frac{3+m}{2m})B^{\frac{3-m}{2m}}\right)\\\nonumber
&&+\frac{C_3}{2\sqrt{T}}-\frac{8b_1\pi}{3h_0^2}-\frac{\sqrt{8\pi\frac{8}{3}}b_2(m+1)\frac{5}{2}T^{\frac{3}{2}}}{15h_0^5}\}
-18h_0^4e^{4mN}(j-3q-4)\left(\frac{9-m^2}{4m^2}\right)C_1B^{\frac{3(1-m)}{2m}}
\\\nonumber&&+36h_0^4e^{4mN}(1+q)-24h_0^4e^{4mN}(1+q)\{-\frac{C_3}{4}T^{-\frac{3}{2}}
-\frac{\sqrt{8\pi\frac{8}{3}}b_2(m+1)T^{\frac{1}{2}}}{4h_0^5}\}+36h_0^4e^{4mN}(j-3q-4)^2\\\label{p2*}
&&\times\{\frac{3(1-m)(9-m^2)}{8m^3}C_1B^{\frac{3-5m}{2m}}\}\}\geq0,\\\nonumber
&&8\pi\left(b_0+b_1e^{2mN}+\frac{96(m+1)}{5}b_2e^{5mN}\right)-18h_0^4e^{4mN}(j-3q-4)
\left(\frac{9-m^2}{4m^2}\right)C_1B^{\frac{3(1-m)}{2m}}\\\nonumber
&&-3h_0^2e^{2mN}(1+q)\left(C_2+C_1B^{\frac{3-m}{2m}}\frac{3+m}{2m}\right)-\frac{1}{2}\{C_1B^{\frac{3+m}{2m}}+C_2B+16\pi
b_0+C_3\sqrt{T}-\frac{8\pi b_1T}{3h_0^2}\\\label{p3*}
&&-\frac{\sqrt{8\pi\frac{8}{3}}b_2(m+1)T^{\frac{5}{2}}}{15h_0^5}\}\geq0.
\end{eqnarray}
\begin{figure}
\center\epsfig{file=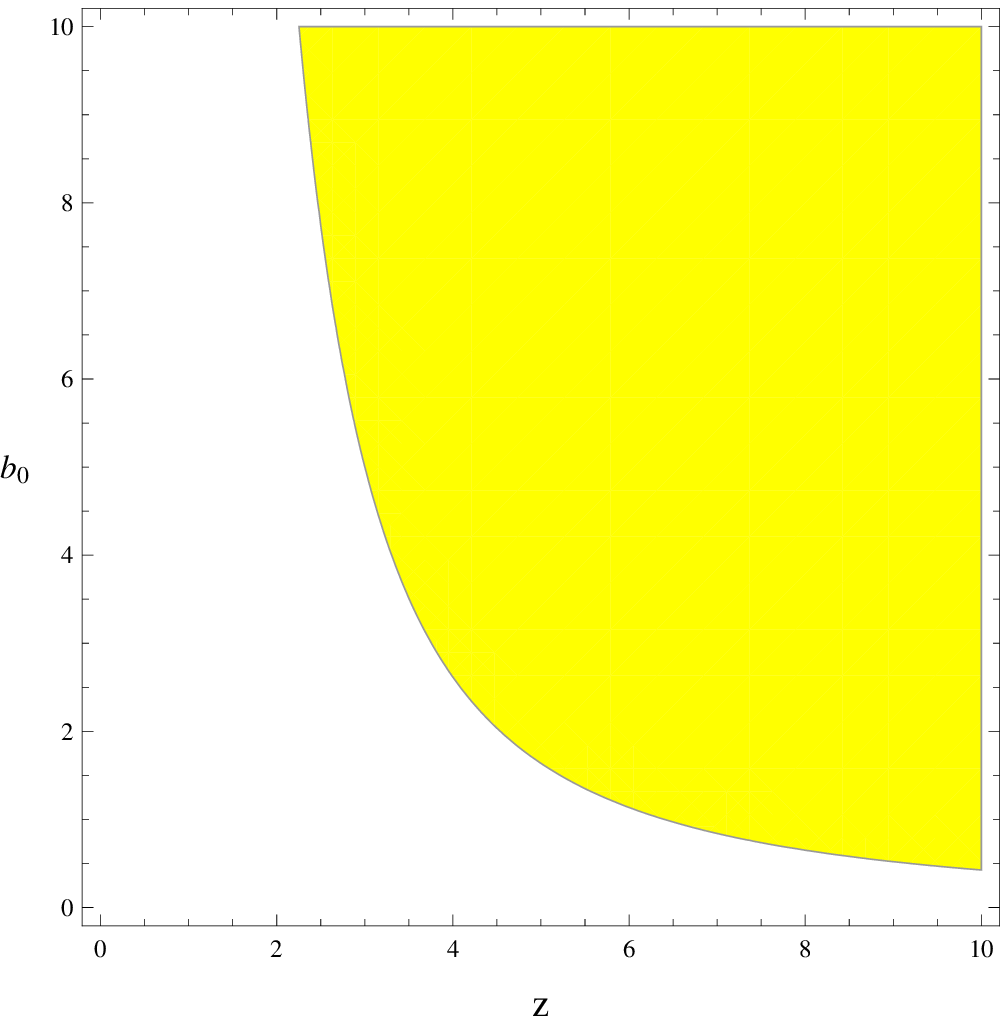, width=0.35\linewidth}
\epsfig{file=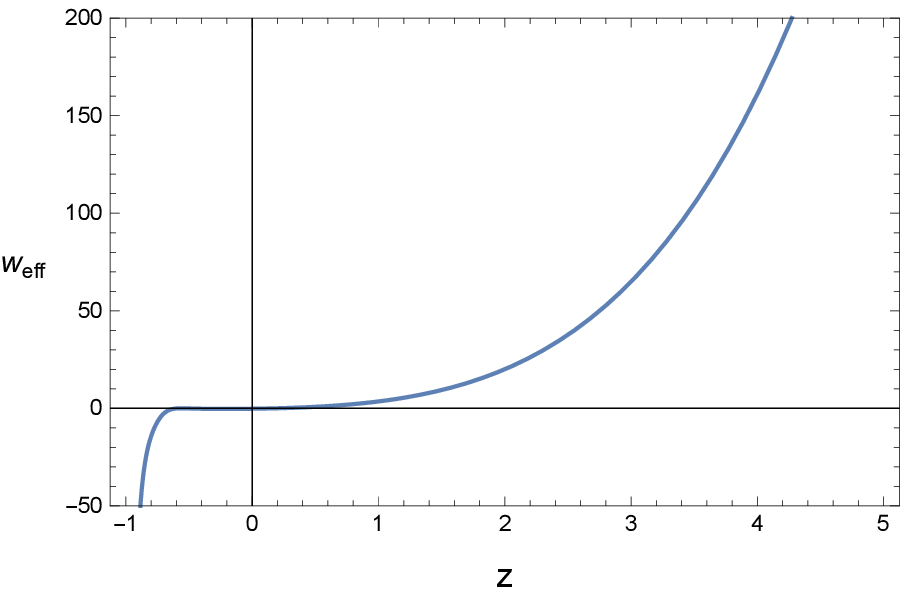, width=0.35\linewidth}\caption{Left region
plot shows validity region for WEC and NEC while Right plot
indicates the evolution of effective EoS parameter versus cosmic
time. For right plot, we assume
$h_0=-0.1,~\omega_m=0,~m=2,~j=1.02,~q=-1,~b_0=1,~b_1=2,~b_2=0.0001$ and
$C_1=C_2=C_3=1.$}
\end{figure}
The graphical illustration of these energy bounds versus red shift
parameter and constant $b_0$ are given in Figure \textbf{4}.
Furthermore, the effective EoS parameter turns out to be
\begin{eqnarray}\nonumber
\omega_{eff}&=&[\omega_m\rho_m+\frac{1}{8\pi}\{\frac{1}{2}(C_1B^{\frac{3+m}{2m}}+C_2B+16\pi
b_0+C_3\sqrt{T}-\frac{8\pi
b_1T}{3h_0^2}-\frac{\sqrt{8\pi\frac{8}{3}}b_2(m+1)T^{\frac{5}{2}}}{15h_0^5})\\\nonumber
&+&mh_0^2e^{2mN}(2(\frac{C_3}{2\sqrt{T}}-\frac{8\sqrt{\frac{8}{3}}\pi
b_2(m+1)}{15h_0^5}\frac{5}{2}T^{\frac{3}{2}})-3(C_1\left(\frac{3+m}{2m}\right)B^{\frac{3+m}{2m}-1}+C_2))\\\nonumber
&-&2h_0e^{mN}(12mh_0^3e^{3mN}\left(-\frac{C_3}{4T^{\frac{3}{2}}}-\frac{1}{6h_0^5}\sqrt{\frac{8}{3}}8\pi
b_2(m+1)\frac{3}{2}T^{\frac{1}{2}}\right))
-9h_0^2e^{2mN}(h_0e^{mN}\\\nonumber
&\times&\left(C_1\frac{9-m^2}{4m^2}B^{\frac{3(1-m)}{2m}}\right)\left(6h_0^2(m+1)e^{2mN}\right))+
C_1\frac{9-m^2}{4m^2}(m+1)6h_0^4e^{mN}\\\nonumber
&\times&\left(\frac{3(1-m)}{2m}B^{\frac{3(1-m)}{2m}-1}(6h_0^2(m+1)e^{4mN})+2me^{2mN}B^{\frac{3(1-m)}{2m}}\right)\}]
[\rho_m+\frac{1}{8\pi}\{-3h_0e^{mN}(h_0e^{mN}
\end{eqnarray}
\begin{eqnarray}\nonumber
&\times&\left(C_1\frac{9-m^2}{4m^2}B^{\frac{3(1-m)}{2m}}\right)\left(6h_0^2(m+1)e^{2mN}\right))
+(3mh_0^2e^{2mN}+9h_0^2e^{2mN})(C_1\left(\frac{3+m}{2m}\right)B^{\frac{3+m}{2m}-1}\\\label{p4*}
&+&C_2)-\frac{1}{2}(C_1B^{\frac{3+m}{2m}}+C_2B+16\pi
b_0+C_3\sqrt{T}-\frac{8\pi
b_1T}{3h_0^2}-\frac{\sqrt{8\pi\frac{8}{3}}b_2(m+1)T^{\frac{5}{2}}}{15h_0^5})\}]^{-1}.
\end{eqnarray}
The possible validity of energy constraints and the dynamics of the
effective EoS parameter is illustrated in the Figure \textbf{4}. In
the left penal, the validity region of NEC and WEC is given for the
described choice of free parameters. It is shown that the energy
constraints are valid for increasing values of $b_0$ and redshift
$z$, in particular, $z\geq3$ (decreasing rate of scale factor). For
very small values of $b_0$ closer to zero, these constraints remain
invalid. The EoS parameter indicates the negative behavior as shown
in the right penal.

\section{Evolution of Energy Bounds using Reconstructed Models for $f(T,B)=-T+F(B)$ cosmology}

In this section, we will study the specific case where the function
takes the form $f(T,B)=-T+F(B)$, which is similar to models of the
form $f(R)=R+F(R)$ and $f(T)=-T+f(T)$ studied in $f(R)$ and $f(T)$
gravity, respectively. This theory is equivalent than to consider a
teleparallel background (or $GR$) plus an additional function which
depends on the boundary term which can be also understood as $F(B)=
F(T + R)$. Using $f(T,B)=-T+F(B)$ in the energy conditions
(\ref{005}) and (\ref{006}), we get this specific form of energy
conditions:
\begin{eqnarray}\nonumber
NEC&:& 36H^{4}(1+q)+\kappa
^{2}(p_{m}+\rho_{m})-2H^{2}(1+q)(-1+3F_{B})-18H^{4}(-4+j-3q)F_{BB}\\\label{1}
&+&36H^{6}(-4+j-3q)^{2}F_{BBB}\geq 0,\\\label{2}
WEC&:&\frac{1}{2}(T-F(B))+\kappa
^{2}\rho_{m}-3H^{2}(1+q)F_{B}-18H^{4}(-4+j-3q)F_{BB}\geq 0.
\end{eqnarray}
Similar to the previous section, now we will check the evolution of
energy condition bounds as well as EoS parameter for previously
mentioned four specific cases.

For a de-Sitter reconstruction, the scale factor behaves as
$a(t)=a_{0}e^{H_{o}t}$, then $B=18H_{0}^{2}$ and hence we consider
$F(B)$ as
\begin{eqnarray}\label{4}
F(B)=C_{1}e^{\frac{B}{18H_{o}^{2}}}-2(3H_{0}^{2}-\kappa^{2}\rho_{0}).
\end{eqnarray}
Here $C_{1}$ is an integration constant. Using (\ref{4}) in
(\ref{1}) and (\ref{2}), we get
\begin{eqnarray}
NEC&:&324H_0^{2}(1+18H_0^{2})(1+q)+e(-2+j^{2}-3q+9q^{2}-j(17+6q))C_{1} \geq 0\\
WEC&:& 108H_0^{2}-e(8+j)C_{1}\geq 0
\end{eqnarray}
These constraints involve parameters $H_0,~q,~j$ and $C_1$. The
consistency of these energy bounds for $q$ parameter and $C_1$ is
shown in left plot of Figure \textbf{5}. Clearly, it indicates that
these energy bounds will remain valid if $C_1$ is negative while
$-1\leq q\leq 2$ that corresponds to quintessence, radiation, matter
and dust dominated cosmic epochs. Thus it can be concluded that in
phantom cosmic phase, these bounds will not be consistent. Also, the
effective EoS parameter, in this case, takes the following form:
\begin{eqnarray}
\omega_{eff}=\frac{\omega_m\rho_0e^{-3(1+\omega_m)H_0t}-\frac{1}{\kappa^2}(6H_0^2+\kappa^2\rho_0)}{\rho_0e^{-3(1+\omega_m)H_0t}
+\frac{1}{\kappa^2}(6H_0^2-\kappa^2\rho_0)}.
\end{eqnarray}
The right plot of Figure \textbf{5} represents the behavior of this
parameter versus cosmic time. Clearly, its negative values less than
-1 corresponds to the phantom stage of cosmic evolution.
\begin{figure}
\center\epsfig{file=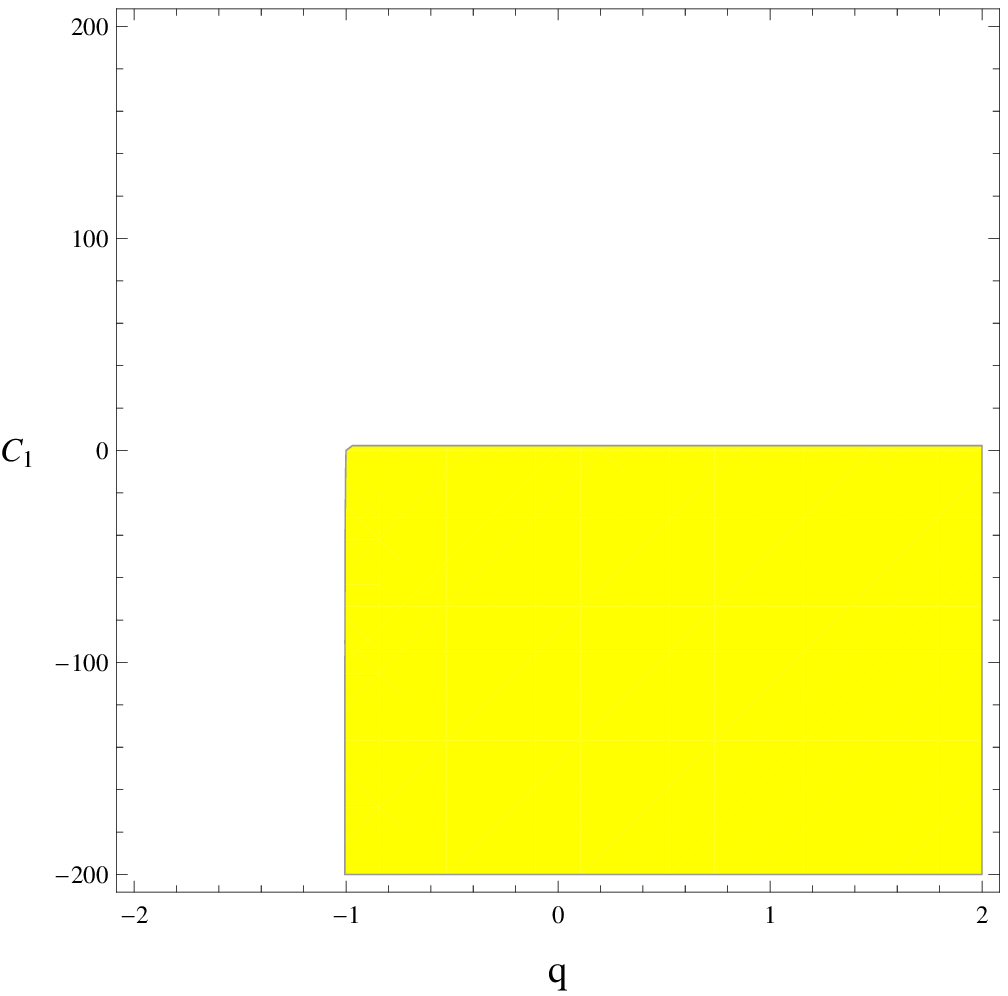, width=0.35\linewidth}
\epsfig{file=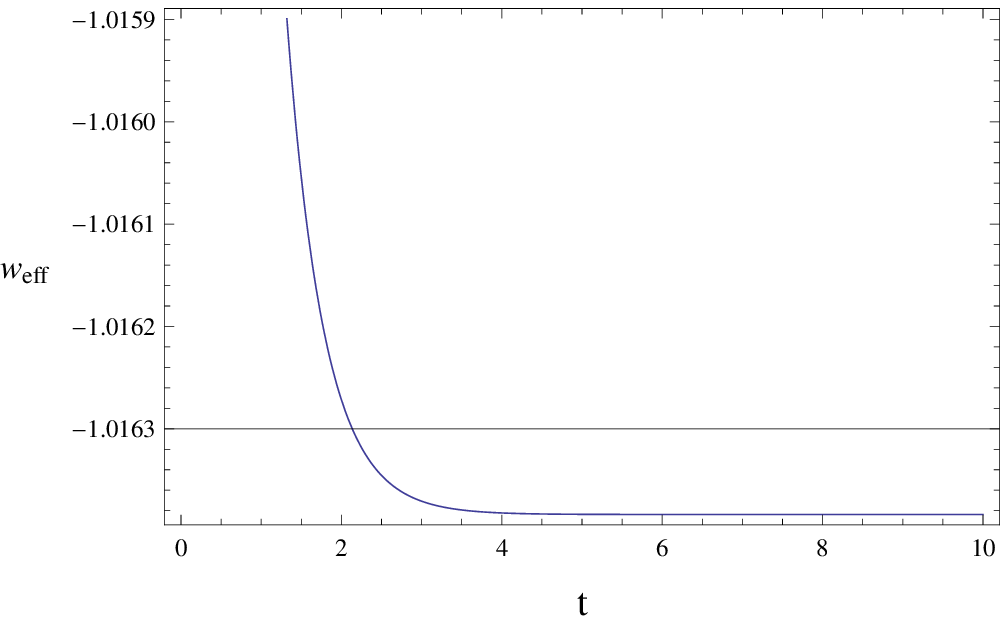, width=0.35\linewidth}\caption{Left region
plot shows validity region for WEC and NEC while Right plot
indicates the evolution of effective EoS parameter versus cosmic
time. For left plot, $H_0=0.718$ and $j=1.02$. For right plot, we
assume $\rho_0=0.001$ and $\omega_m=0$.}
\end{figure}

For a power-law cosmology, we take $F(B)$ in the following form:
\begin{eqnarray}\nonumber
F(B)&=&C_{1}B^{\frac{1-3h}{2}}+B\left(C_{2}-\frac{2h}{(3h+1)^{2}}
\right)+\frac{hBLog(B)}{3h+1}+\frac{BhLog(9h+3)}{3h+1}\\\label{3}&-&\frac{\kappa
^{2}\rho _{0}\left((3h-1)^{1-\frac{3}{2}h(\omega_m+1)}\times
2^{2-\frac{3}{2}h(\omega_m+1)}\right)\left(\frac{Bt_{0}^{2}}{3h}
\right)^{\frac{3}{2}h(\omega_m+1)}}{\left(3h(\omega_m+1)-2
\right)\left(3h(\omega_m+2)-1\right)},
\end{eqnarray}
where $C_{1}$ and $C_{2}$ are integration constants. Using (\ref{3})
in (\ref{1}) and (\ref{2}), we get
\begin{eqnarray}\nonumber
NEC&:&36H^{4}(1+q)+\frac{H^{6}(-4+j-3q)^{2}t^{4}}{12h^{3}(-1+3h)^{3}}\left(-\frac{12h^{2}(-1+3h)}{1+3h}
-2^{-\frac{3}{2}(1+h)}\times 3^{-\frac{3}{2}(-1+h)}(1+h)(9h^{2}-1)\right.\\
&&\left. \nonumber \times
\left(\frac{h(-1+3h)}{t^{2}}\right)^{\frac{1}{2}(1-3h)}t^{2}C_{1}\right)
-\frac{H^{4}(-4+j-3q)t^{2}\left(\frac{12(-1+3h)}{1+3h}+\frac{6^{\frac{1}{2}-\frac{3h}{2}}
(-1+\frac{1}{2}(1-3h))(1-3h)\left( \frac{h(-1+3h)}{t^{2}}
\right)^{\frac{1}{2}
-\frac{3h}{2}}t^{2}C_{1}}{h^{2}}\right)}{4(1-3h)^{2}}\\ \nonumber &&
-2H^{2}(1+q)\Bigg[-1+ \frac{3}{2}\Bigg[-\frac{6^{-\frac{1}{2}
-\frac{3h}{2}}\left(\frac{h(-1+3h)}{t^{2}}\right)^{\frac{1}{2}(1-3h)}t^{2}C_{1}}{h}\\&&
+\frac{2(h(-1+3h+ (1+3h)Log(3+9h)+(1+3h)Log(\frac{6h(-1+3h)}{t^{2}}))+(1+3h)^{2}C_{2})}{(1+3h)^{2}}\Bigg]\Bigg]\geq 0 \\
WEC&:&\nonumber
\frac{1}{ht^{2}}\Bigg[-2^{-\frac{1}{2}-\frac{3h}{2}}\times
3^{\frac{1}{2}-\frac{3h}{2}}\left(\frac{h(-1+3h)}{t^{2}}
\right)^{\frac{1}{2}(-1-3h)}\left(4h^{2}(-1+3h)-2h(-1+3h)H^{2}(1+q)t^{2}
\right. \\&& \nonumber
\left.+(1+3h)H^{4}(-4+j-3q)t^{4}\right)C_{1}\Bigg]+
\frac{12}{(-1+3h)(t+3ht)^{2}}\Bigg[h^{2}(-1+3h)(-1+3h(4+3h))\\ &&
\nonumber
-(1-3h)^{2}hH^{2}(1+q)t^{2}+(1+3h)H^{4}(4-j+3q)t^{4}-h(-1+9h^{2})(h(-1+3h)+H^{2}(1+q)t^{2})\\
&& \left(Log(3+9h)+Log\left(\frac{6h(-1+3h)}{t^{2}}
\right)-(-1+3h)(1+3h)^{2}\left(h(-1+3h)+H^{2}(1+q)t^{2}\right)C_{2}
\right)\Bigg]\geq 0
\end{eqnarray}
Also, effective EoS parameter can be written as
\begin{eqnarray}\nonumber
\omega_{eff}&=&[\omega_m\rho_m+\frac{1}{8\pi}\{\frac{1}{2}(-T+C_1B^{\frac{1-3h}{2}}+B\left(C_2-\frac{2h}{(3h+1)^2}\right)
+\frac{h}{3h+1}B\log(B)+\frac{h\log(9h+3)}{3h+1}B\\\nonumber
&-&\frac{8\pi\left((3h-1)^{1-\frac{3}{2}h(\omega_m+1)}2^{2-\frac{3}{2}h(\omega_m+1)}\right)}{(3h(\omega_m+1)-2)(3h(\omega_m+2)-1)}
(\frac{Bt_0^2}{3h})^{\frac{3}{2}h(\omega_m+1)})+\frac{h}{t^2}(2+3(C_1\frac{1-3h}{2}B^{\frac{1-3h}{2}-1}\\\nonumber
&+&\left(C_2-\frac{2h}{(3h+1)^2}\right)-\frac{h}{3h+1}\left(\log(B)+\frac{1}{\ln(10)}\right)
+\frac{h\log(9h+3)}{3h+1}
\end{eqnarray}
\begin{eqnarray}\nonumber
&-&\frac{8\pi\left((3h-1)^{1-\frac{3}{2}h(\omega_m+1)}2^{2-\frac{3}{2}h(\omega_m+1)}\right)}{(3h(\omega_m+1)-2)(3h(\omega_m+2)-1)}
(\frac{t_0^2}{3h})^{\frac{3}{2}h(\omega_m+1)}(\frac{3}{2}h(\omega_m+1))B^{\frac{3}{2}h(\omega_m+1)-1}))\\\nonumber
&-&9(\frac{h}{t})^2(C_1\frac{1-3h}{2}B^{\frac{1-3h}{2}-1}+\left(C_2-\frac{2h}{(3h+1)^2}\right)-\frac{h}{3h+1}\left(\log(B)+\frac{1}{\ln(10)}\right)
+\frac{h\log(9h+3)}{3h+1}
\\\nonumber
&-&\frac{8\pi\left((3h-1)^{1-\frac{3}{2}h(\omega_m+1)}2^{2-\frac{3}{2}h(\omega_m+1)}\right)}{(3h(\omega_m+1)-2)(3h(\omega_m+2)-1)}
(\frac{t_0^2}{3h})^{\frac{3}{2}h(\omega_m+1)}(\frac{3}{2}h(\omega_m+1))B^{\frac{3}{2}h(\omega_m+1)-1})
\\\nonumber
&+&\left(-\frac{12h(3h-1)}{t^3}\right)^2\{C_1\frac{9h^2-1}{4}(\frac{1-3h}{2}-2)B^{\frac{1-3h}{2}-3}+\frac{h}{3h+1}\left(\frac{1}{B^2\ln(10)}\right)
\\\nonumber
&-&\frac{8\pi\left((3h-1)^{1-\frac{3}{2}h(\omega_m+1)}2^{2-\frac{3}{2}h(\omega_m+1)}\right)}{(3h(\omega_m+1)-2)(3h(\omega_m+2)-1)}
(\frac{t_0^2}{3h})^{\frac{3}{2}h(\omega_m+1)}(\frac{3}{2}h(\omega_m+1))(\frac{3}{2}h(\omega_m+1)-1)\\\nonumber
&\times&(\frac{3}{2}h(\omega_m+1)-2)
B^{\frac{3}{2}h(\omega_m+1)-3}\}+\{C_1\frac{9h^2-1}{4}B^{\frac{1-3h}{2}-2}-\frac{h}{3h+1}\left(\frac{1}{B\ln(10)}\right)\\\nonumber
&-&\frac{8\pi\left((3h-1)^{1-\frac{3}{2}h(\omega_m+1)}2^{2-\frac{3}{2}h(\omega_m+1)}\right)}{(3h(\omega_m+1)-2)(3h(\omega_m+2)-1)}
(\frac{t_0^2}{3h})^{\frac{3}{2}h(\omega_m+1)}(\frac{3}{2}h(\omega_m+1))(\frac{3}{2}h(\omega_m+1)-1)\\\nonumber
&\times&B^{\frac{3}{2}h(\omega_m+1)-2}\}\left(\frac{36h(3h-1)}{t^4}\right)\}]
[\rho_m+\frac{1}{8\pi}\{-3\frac{h}{t}(\left(-\frac{12h(3h-1)}{t^3}\right)\{C_1\frac{9h^2-1}{4}B^{\frac{1-3h}{2}-2}\\\nonumber
&-&\frac{h}{3h+1}\left(\frac{1}{B\ln(10)}\right)-\frac{8\pi\left((3h-1)^{1-\frac{3}{2}h(\omega_m+1)}2^{2-\frac{3}{2}h(\omega_m+1)}\right)}{(3h(\omega_m+1)-2)(3h(\omega_m+2)-1)}
(\frac{t_0^2}{3h})^{\frac{3}{2}h(\omega_m+1)}(\frac{3}{2}h(\omega_m+1))\\\nonumber
&\times&(\frac{3}{2}h(\omega_m+1)-1)B^{\frac{3}{2}h(\omega_m+1)-2}\})
+\left(\frac{9h^2}{t^2}-\frac{3h}{t^2}\right)(\left(-\frac{12h(3h-1)}{t^3}\right)\{C_1\frac{9h^2-1}{4}B^{\frac{1-3h}{2}-2}\\\nonumber
&-&\frac{h}{3h+1}\left(\frac{1}{B\ln(10)}\right)-\frac{8\pi\left((3h-1)^{1-\frac{3}{2}h(\omega_m+1)}2^{2-\frac{3}{2}h(\omega_m+1)}\right)}{(3h(\omega_m+1)-2)(3h(\omega_m+2)-1)}
(\frac{t_0^2}{3h})^{\frac{3}{2}h(\omega_m+1)}(\frac{3}{2}h(\omega_m+1))\\\nonumber
&\times&(\frac{3}{2}h(\omega_m+1)-1)B^{\frac{3}{2}h(\omega_m+1)-2}\})-\frac{1}{2}
(-T+C_1\frac{1-3h}{2}B^{\frac{1-3h}{2}-1}+\left(C_2-\frac{2h}{(3h+1)^2}\right)\\\nonumber
&-&\frac{h}{3h+1}\left(\log(B)+\frac{1}{\ln(10)}\right)+\frac{h\log(9h+3)}{3h+1}
-\frac{8\pi\left((3h-1)^{1-\frac{3}{2}h(\omega_m+1)}2^{2-\frac{3}{2}h(\omega_m+1)}\right)}{(3h(\omega_m+1)-2)(3h(\omega_m+2)-1)}\\\nonumber
&\times&(\frac{t_0^2}{3h})^{\frac{3}{2}h(\omega_m+1)}(\frac{3}{2}h(\omega_m+1))B^{\frac{3}{2}h(\omega_m+1)-1})\}]^{-1}.
\end{eqnarray}
\begin{figure}
\center\epsfig{file=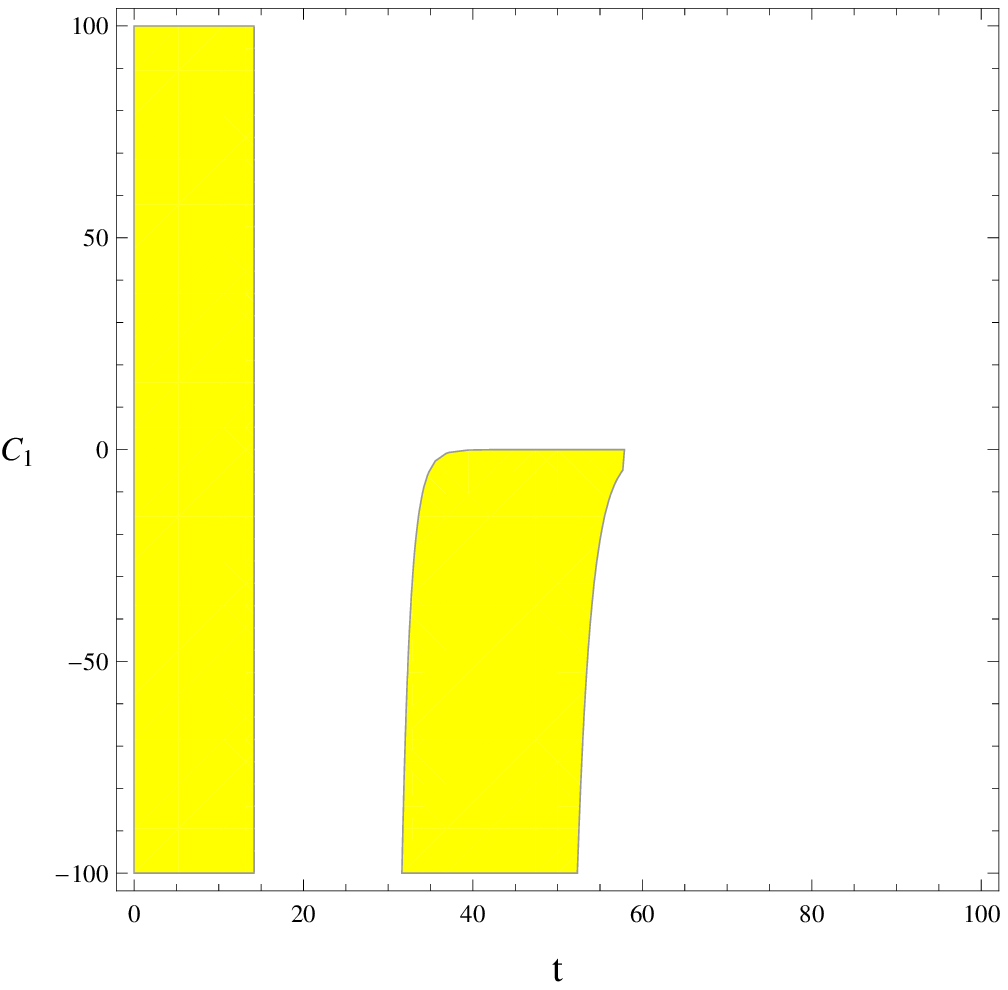, width=0.35\linewidth}
\epsfig{file=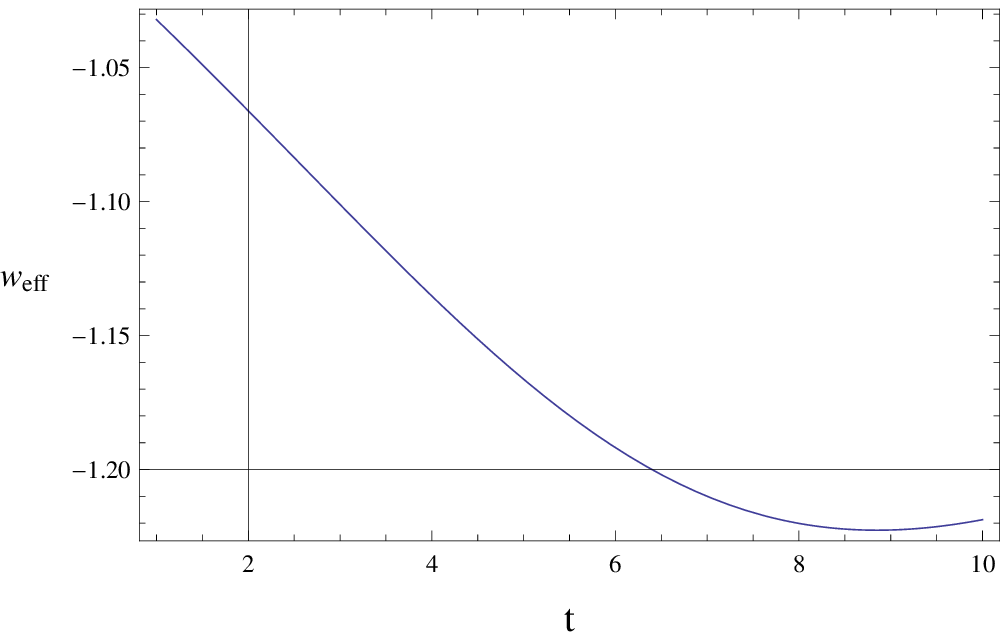, width=0.35\linewidth}\caption{Left region
plot shows validity region for WEC and NEC while Right plot
indicates the evolution of effective EoS parameter versus cosmic
time. Here $\rho_0=t_0=C_2=C_3=1$ and $\omega_m=0,~h=10$. For right
plot, $C_1=10$.}
\end{figure}
The graphical illustration for energy constraints and effective EoS
parameter is presented in the Figure \textbf{6}. The left penal
provides the possible validity region for energy bounds. It is found
that for initial cosmic time, these constraints are valid with
$-100\leq C_1\leq100$. Further, these constraints are also valid if
$C_1\leq0$ and $30\leq t\leq50$. The right penal indicates the
dynamics of EoS parameter versus cosmic time. Its negative behavior
less than -1 corresponds to phantom accelerated phase of cosmos.

For $\Lambda CDM$ Universe, we choose $F(B)$ as
\begin{eqnarray} \label{5}\nonumber
&&F(B)=BC_{1}+\frac{log(B-18H_{0}^{2})(6a_{0}^{3}Bl-2B\kappa
^{2}\rho _{0})+6a_{0}^{3}l(B-9H_{0}^{2})+\kappa ^{2}\rho
_{0}(B-36H_{0}^{2})}{27a_{0}^{3}l} \\ && +
\frac{C_{2}\left(3H_{0}\sqrt{B-9H_{0}^{2}}-B\times
arctan\left(\frac{\sqrt{B-9H_{0}^{2}}}{3H_{0}}\right)\right)}{54H_{0}^{3}}.
\end{eqnarray}
The corresponding weak and null energy condition bounds are given by
\begin{eqnarray}\nonumber
&&36(\sqrt{H_0^2+le^{-3N}})^4(1+q)+8\pi\rho_0(1+\omega_m)\left(\frac{a_0}{1+z}\right)^{-3(\omega_m+1)}-2(H_0^2+le^{-3N})(1+q)
(-1+3\{C_1\\\nonumber
&&+\frac{1}{27a_0^3l}\left(\frac{B(6a_0^3l-16\pi\rho_0)}{\ln(10)(B-18H_0^2)}+\log(B-18H_0^2)(6a_0^3l-16\pi\rho_0)+
6a_0^3l+8\pi\rho_0\right)-\frac{C_2}{54H_0^3}\arctan\left(\frac{\sqrt{B-9H_0^2}}{3H_0}\right)\})\\\nonumber
&&-18H_0^2(H_0^2+le^{-3N})^2(-4+j-3q)(\frac{1}{27a_0^3l}\left(\frac{(6a_0^3l-16\pi\rho_0)}{\ln(10)}\left(\frac{1}{B-18H_0^2}
-\frac{B}{(B-18H_0^2)^2}\right)
+\frac{(6a_0^3l-16\pi\rho_0)}{\ln(10)(B-18H_0^2)}\right)\\\nonumber
&&-\frac{3C_2}{108H_0^2}B^{-1}(B-9H_0^2)^{-\frac{1}{2}})+36(H_0^2+le^{-3N})(-4+j-3q)^2(\frac{1}{27a_0^3l}\\\nonumber
&\times&\left(\frac{(6a_0^3l-16\pi\rho_0)}{\ln(10)}\left(-\frac{2}{(B-18H_0^2)^2}+\frac{2B}{(B-18H_0^2)^3}\right)
-\frac{(6a_0^3l-16\pi\rho_0)}{\ln(10)(B-18H_0^2)^2}\right)\\\nonumber
&+&\frac{C_2}{12H_0}\left(B^{-2}(B-9H_0^2)^{-\frac{1}{2}}+B^{-1}\frac{1}{2}(B-9H_0^2)^{-\frac{3}{2}}\right))\geq0,\\\nonumber
&&\frac{1}{2}\{6le^{-3N}+6H_0^2-(BC_{1}+\frac{log(B-18H_{0}^{2})(6a_{0}^{3}Bl-2B\kappa
^{2}\rho _{0})+6a_{0}^{3}l(B-9H_{0}^{2})+\kappa
^{2}\rho_{0}(B-36H_{0}^{2})}{27a_{0}^{3}l}\\\nonumber &&
+\frac{C_{2}\left(3H_{0}\sqrt{B-9H_{0}^{2}}-B\times
arctan\left(\frac{\sqrt{B-9H_{0}^{2}}}{3H_{0}}\right)\right)}{54H_{0}^{3}})\}+8\pi\rho_0\left(\frac{a_0}{1+z}\right)^{-3(1+\omega_m)}\\\nonumber
&&-3(H_0^2+le^{-3N})(1+q)(C_1+\frac{1}{27a_0^3l}\left(\frac{B(6a_0^3l-16\pi\rho_0)}{\ln(10)(B-18H_0^2)}+\log(B-18H_0^2)(6a_0^3l-16\pi\rho_0)+
6a_0^3l+8\pi\rho_0\right)\\\nonumber
&&-\frac{C_2}{54H_0^3}\arctan\left(\frac{\sqrt{B-9H_0^2}}{3H_0}\right))-18(H_0^2+le^{-3N})^2(-4+j-3q)\\\nonumber
&&\times(\frac{1}{27a_0^3l}\left(\frac{(6a_0^3l-16\pi\rho_0)}{\ln(10)}\left(\frac{1}{B-18H_0^2}-\frac{B}{(B-18H_0^2)^2}\right)
+\frac{(6a_0^3l-16\pi\rho_0)}{\ln(10)(B-18H_0^2)}\right)\\\label{l2*}
&&-\frac{3C_2}{108H_0^2}B^{-1}(B-9H_0^2)^{-\frac{1}{2}})\geq0.
\end{eqnarray}
The corresponding effective EoS parameter is given by
\begin{eqnarray}\nonumber
\omega_{eff}&=&[\omega_m\rho_m+\frac{1}{8\pi}\{\frac{1}{2}(-T+BC_{1}+\frac{log(B-18H_{0}^{2})(6a_{0}^{3}Bl-16\pi
B\rho_{0})+6a_{0}^{3}l(B-9H_{0}^{2})+8\pi\rho_{0}(B-36H_{0}^{2})}{27a_{0}^{3}l}\\\nonumber
&+&\frac{C_{2}\left(3H_{0}\sqrt{B-9H_{0}^{2}}-B\times
arctan\left(\frac{\sqrt{B-9H_{0}^{2}}}{3H_{0}}\right)\right)}{54H_{0}^{3}})+\frac{3}{2}le^{-3N}(2
+3(C_1\\\nonumber
&+&\frac{1}{27a_0^3l}\left(\frac{B(6a_0^3l-16\pi\rho_0)}{\ln(10)(B-18H_0^2)}+\log(B-18H_0^2)(6a_0^3l-16\pi\rho_0)+
6a_0^3l+8\pi\rho_0\right)\\\nonumber
&-&\frac{C_2}{54H_0^3}\arctan\left(\frac{\sqrt{B-9H_0^2}}{3H_0}\right)))-9H^2(C_1+\frac{1}{27a_0^3l}\{\frac{B(6a_0^3l
-16\pi\rho_0)}{\ln(10)(B-18H_0^2)}+\log(B-18H_0^2)(6a_0^3l\\\nonumber
&-&16\pi\rho_0)+6a_0^3l+8\pi\rho_0\}-\frac{C_2}{54H_0^3}\arctan\left(\frac{\sqrt{B-9H_0^2}}{3H_0}\right))+(H_0^2+le^{-3N})^{\frac{1}{2}}
((-27le^{-3N})^2(\frac{1}{27a_0^3l}\\\nonumber
&\times&\left(\frac{(6a_0^3l-16\pi\rho_0)}{\ln(10)}\left(-\frac{2}{(B-18H_0^2)^2}+\frac{2B}{(B-18H_0^2)^3}\right)
-\frac{(6a_0^3l-16\pi\rho_0)}{\ln(10)(B-18H_0^2)^2}\right)\\\nonumber
&+&\frac{C_2}{12H_0}\left(B^{-2}(B-9H_0^2)^{-\frac{1}{2}}+B^{-1}\frac{1}{2}(B-9H_0^2)^{-\frac{3}{2}}\right))
+(81le^{-3N})(\frac{1}{27a_0^3l}\{\frac{(6a_0^3l-16\pi\rho_0)}{\ln(10)}\\\nonumber
&\times&\left(\frac{1}{B-18H_0^2}-\frac{B}{(B-18H_0^2)^2}\right)+\frac{(6a_0^3l-16\pi\rho_0)}{\ln(10)(B-18H_0^2)}\}
-\frac{3C_2}{108H_0^2}B^{-1}(B-9H_0^2)^{-\frac{1}{2}}))\}]\\\nonumber
&\times&[\rho_m+\frac{1}{8\pi}\{3(H_0^2+le^{-3N})(27le^{-3N})(\frac{1}{27a_0^3l}
\{\frac{(6a_0^3l-16\pi\rho_0)}{\ln(10)}\left(\frac{1}{B-18H_0^2}
-\frac{B}{(B-18H_0^2)^2}\right)\\\nonumber
&+&\frac{(6a_0^3l-16\pi\rho_0)}{\ln(10)(B-18H_0^2)}\}-\frac{3C_2}{108H_0^2}B^{-1}(B-9H_0^2)^{-\frac{1}{2}})+
(-\frac{9}{2}le^{-3N}+9(H_0^2+le^{-3N}))(C_1\\\nonumber
&+&\frac{1}{27a_0^3l}\left(\frac{B(6a_0^3l-16\pi\rho_0)}{\ln(10)(B-18H_0^2)}+\log(B-18H_0^2)(6a_0^3l-16\pi\rho_0)+
6a_0^3l+8\pi\rho_0\right)\\\nonumber
&-&\frac{C_2}{54H_0^3}\arctan\left(\frac{\sqrt{B-9H_0^2}}{3H_0}\right))-\frac{1}{2}(-T
+BC_{1}\\\nonumber
&+&\frac{log(B-18H_{0}^{2})(6a_{0}^{3}Bl-16B\pi\rho_{0})+6a_{0}^{3}l(B-9H_{0}^{2})+8\pi\rho_{0}(B-36H_{0}^{2})}{27a_{0}^{3}l}
\\\label{l3*}
&+&\frac{C_{2}\left(3H_{0}\sqrt{B-9H_{0}^{2}}-B\times
arctan\left(\frac{\sqrt{B-9H_{0}^{2}}}{3H_{0}}\right)\right)}{54H_{0}^{3}})\}]^{-1}.
\end{eqnarray}
\begin{figure}
\center\epsfig{file=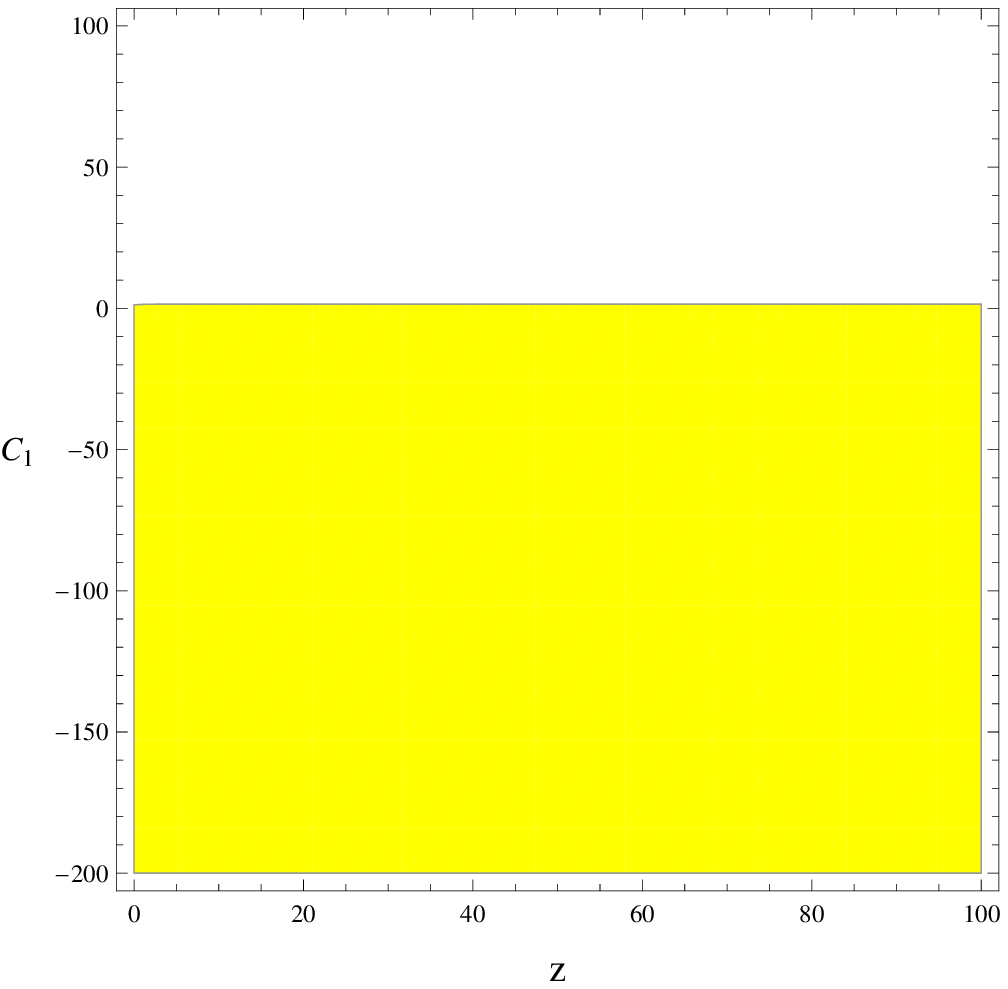, width=0.35\linewidth}
\epsfig{file=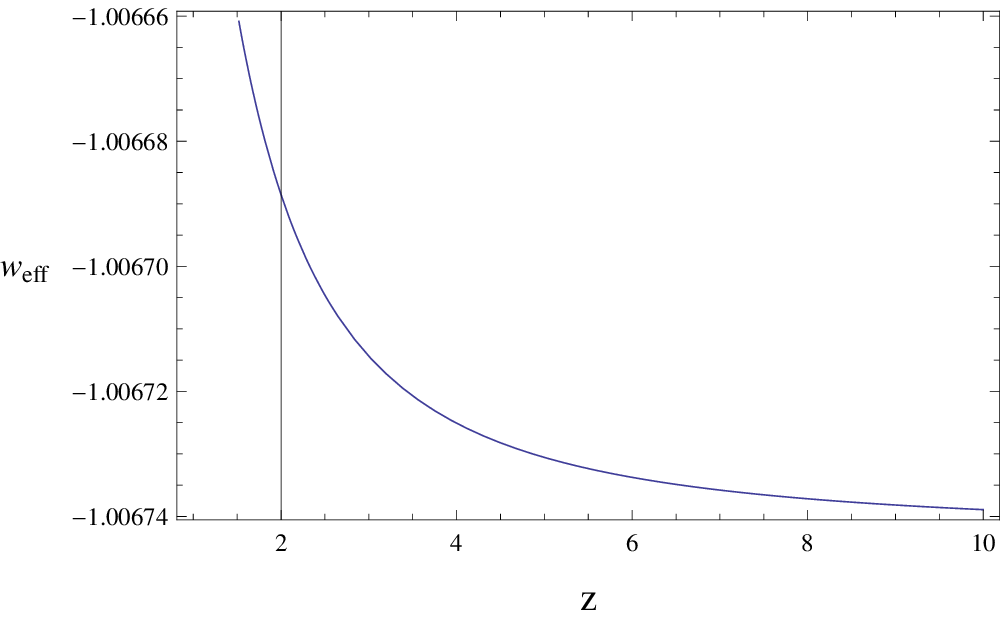, width=0.35\linewidth}\caption{Left region
plot shows validity region for WEC and NEC while Right plot
indicates the evolution of effective EoS parameter versus cosmic
time. Here $\rho_0=t_0=C_2=1,~C_1=100,~C_3=-0.1$ and
$\omega_m=0,~l=\frac{8\pi}{3}$. For right plot, we consider $C_2=10$
and $q=-0.64$.}
\end{figure}
Figure \textbf{7} provides the graphical interpretation of these
constraints as well as the dynamics of effective EoS parameter for
this model. Clearly the left penal indicates that these constraints
are valid with increasing redshift parameter where $C_1<0$. The
right penal shows the negative values of this parameter, i.e.,
$\omega_{eff}<-1$, thus leading to the phantom stage of cosmic
evolution.

The reconstructed phantom cosmological model is given by
\begin{eqnarray}\nonumber
F(B)&=&-\frac{(m-3)B\log(B)(3h_0^2-8\pi
b_1)}{3h_0^2(m-3)^2}-\frac{16b_1m\pi-3h_0^2(C_2m^2-6C_2m+9C_2+4m)B}{3h_0^2(m-3)^2}\\\label{p6*}
&+&\frac{16\sqrt{0.67}8b_2\pi(m+1)}{45h_0^5(m+3)^{3/2}(4m-3)}B^{\frac{5}{2}}+C_1B^{\frac{m+3}{2m}}+16b_0\pi.
\end{eqnarray}
The energy constraints for this model take the following form:
\begin{eqnarray}\nonumber
&&36(h_0e^{mN})^4(1+q)+8\pi(1+\omega_m)(b_0+b_1e^{mN}+\frac{96(m+1)}{5}b_2e^{5mN})-2(h_0^2e^{2mN})(1+q)(-1\\\nonumber
&&+3\{-\frac{(m-3)(3h_0^2-8b_1\pi)}{3h_0^2(m-3)^2}(\log(B)+\frac{1}{\ln(10)})
-\frac{16b_1m\pi-3h_0^2(C_2m^2-6C_2m+9C_2+4m)}{3h_0^2(m-3)^2}
\\\nonumber
&&+\frac{16\sqrt{0.67}8b_2\pi(m+1)}{45h_0^5(m+3)^{3/2}(4m-3)}(\frac{5}{2})B^{\frac{3}{2}}+C_1(\frac{m+3}{2m})B^{\frac{m+3}{2m}-1}\})
-18(h_0e^{mN})^4(-4+j-3q)\\\nonumber
&&\times(-\frac{(m-3)(3h_0^2-8b_1\pi)}{3h_0^2(m-3)^2}(\frac{1}{B\ln(10)})
+\frac{4\sqrt{0.67}8b_2\pi(m+1)}{3h_0^5(m+3)^{3/2}(4m-3)}B^{\frac{1}{2}}+C_1(\frac{m+3}{2m})(\frac{m+3}{2m}-1)B^{\frac{m+3}{2m}-2})
\\\nonumber
&&+36(h_0e^{2mN})^4(-4+j-3q)^2(-\frac{(m-3)(3h_0^2-8b_1\pi)}{3h_0^2(m-3)^2}(\frac{1}{B\ln(10)})
+\frac{\sqrt{\frac{48}{27}}8b_2\pi(m+1)}{h_0^5(m+3)^{3/2}(4m-3)}\frac{1}{2}B^{\frac{-1}{2}}
\\\label{p7*}
&&+C_1(\frac{9-m^2}{4m^2})(\frac{3(1-m)}{2m})B^{\frac{3-5m}{2m}})\geq0,\\\nonumber
&&\frac{1}{2}(6h_0^2e^{2mN}-(-\frac{(m-3)B\log(B)(3h_0^2-8\pi
b_1)}{3h_0^2(m-3)^2}-\frac{16b_1m\pi-3h_0^2(C_2m^2-6C_2m+9C_2+4m)B}{3h_0^2(m-3)^2}
\\\nonumber
&&+\frac{16\sqrt{0.67}8b_2\pi(m+1)}{45h_0^5(m+3)^{3/2}(4m-3)}B^{\frac{5}{2}}+C_1B^{\frac{m+3}{2m}}+16b_0\pi))
+8\pi(b_0+b_1e^{mN}+\frac{96(m+1)}{5}b_2e^{5mN})\\\nonumber
&&-3(h_0e^{mN})^2(1+q)(-\frac{(m-3)(3h_0^2-8b_1\pi)}{3h_0^2(m-3)^2}(\log(B)+\frac{1}{\ln(10)})
-\frac{16b_1m\pi-3h_0^2(C_2m^2-6C_2m+9C_2+4m)}{3h_0^2(m-3)^2}
\\\nonumber
&&+\frac{16\sqrt{0.67}8b_2\pi(m+1)}{45h_0^5(m+3)^{3/2}(4m-3)}(\frac{5}{2})B^{\frac{3}{2}}+C_1(\frac{m+3}{2m})B^{\frac{m+3}{2m}-1})
-18(h_0e^{mN})^4(-4+j-3q)\\\nonumber
&&\times(-\frac{(m-3)(3h_0^2-8b_1\pi)}{3h_0^2(m-3)^2}(\frac{1}{B\ln(10)})
+\frac{4\sqrt{0.67}8b_2\pi(m+1)}{3h_0^5(m+3)^{3/2}(4m-3)}B^{\frac{1}{2}}
+C_1(\frac{m+3}{2m})(\frac{m+3}{2m}-1)B^{\frac{m+3}{2m}-2})\geq0.\\\label{p8*}
\end{eqnarray}
Further, the corresponding effective EoS parameter can be written as
\begin{eqnarray}\nonumber
\omega_{eff}&=&[\omega_m(b_0+b_1e^{mN}+\frac{96(m+1)}{5}b_2e^{5mN})
+\frac{1}{8\pi}\{\frac{1}{2}(-6h_0^2e^{2mN}-\frac{(m-3)B\log(B)(3h_0^2-8\pi
b_1)}{3h_0^2(m-3)^2}\\\nonumber
&-&\frac{16b_1m\pi-3h_0^2(C_2m^2-6C_2m+9C_2+4m)B}{3h_0^2(m-3)^2}
+\frac{16\sqrt{0.67}8b_2\pi(m+1)}{45h_0^5(m+3)^{3/2}(4m-3)}B^{\frac{5}{2}}+C_1B^{\frac{m+3}{2m}}+16b_0\pi)
\end{eqnarray}
\begin{eqnarray}\nonumber
&-&mh_0^2e^{2mN}(2+3(-\frac{(m-3)(3h_0^2-8b_1\pi)}{3h_0^2(m-3)^2}(\log(B)+\frac{1}{\ln(10)})
-\frac{16b_1m\pi-3h_0^2(C_2m^2-6C_2m+9C_2+4m)}{3h_0^2(m-3)^2}\\\nonumber
&+&\frac{16\sqrt{0.67}8b_2\pi(m+1)}{45h_0^5(m+3)^{3/2}(4m-3)}(\frac{5}{2})B^{\frac{3}{2}}
+C_1(\frac{m+3}{2m})B^{\frac{m+3}{2m}-1}))-9h_0^2e^{2mN}(-\frac{(m-3)(3h_0^2-8b_1\pi)}{3h_0^2(m-3)^2}\\\nonumber
&\times&(\log(B)+\frac{1}{\ln(10)})-\frac{16b_1m\pi-3h_0^2(C_2m^2-6C_2m+9C_2+4m)}{3h_0^2(m-3)^2}
+\frac{16\sqrt{0.67}8b_2\pi(m+1)}{45h_0^5(m+3)^{3/2}(4m-3)}(\frac{5}{2})B^{\frac{3}{2}}\\\nonumber
&+&C_1(\frac{m+3}{2m})B^{\frac{m+3}{2m}-1})
+12m(m+3)h_0^3e^{3mN}((12h_0^2m(m+3)e^{2mN})(-\frac{(m-3)(3h_0^2-8b_1\pi)}{3h_0^2(m-3)^2}(\frac{1}{B\ln(10)})
\\\nonumber
&+&\frac{\sqrt{\frac{48}{27}}8b_2\pi(m+1)}{h_0^5(m+3)^{3/2}(4m-3)}\frac{1}{2}B^{\frac{-1}{2}}
+C_1(\frac{9-m^2}{4m^2})(\frac{3(1-m)}{2m})B^{\frac{3-5m}{2m}})
+2m(-\frac{(m-3)(3h_0^2-8b_1\pi)}{3h_0^2(m-3)^2}(\frac{1}{B\ln(10)})
\\\nonumber
&+&\frac{4\sqrt{0.67}8b_2\pi(m+1)}{3h_0^5(m+3)^{3/2}(4m-3)}B^{\frac{1}{2}}
+C_1(\frac{m+3}{2m})(\frac{m+3}{2m}-1)B^{\frac{m+3}{2m}-2}))\}]
[b_0+b_1e^{mN}+\frac{96(m+1)}{5}b_2e^{5mN}\\\nonumber
&+&\frac{1}{8\pi}\{-3h_0e^{mN}(12h_0^3m(m+3)e^{3mN}
(-\frac{(m-3)(3h_0^2-8b_1\pi)}{3h_0^2(m-3)^2}(\frac{1}{B\ln(10)})
+\frac{4\sqrt{0.67}8b_2\pi(m+1)}{3h_0^5(m+3)^{3/2}(4m-3)}B^{\frac{1}{2}}\\\nonumber
&+&C_1(\frac{m+3}{2m})(\frac{m+3}{2m}-1)B^{\frac{m+3}{2m}-2}))
+(3mh_0^2e^{2mN}+9h_0^2e^{2mN})(-\frac{(m-3)(3h_0^2-8b_1\pi)}{3h_0^2(m-3)^2}(\log(B)+\frac{1}{\ln(10)})\\\nonumber
&-&\frac{16b_1m\pi-3h_0^2(C_2m^2-6C_2m+9C_2+4m)}{3h_0^2(m-3)^2}
+\frac{16\sqrt{0.67}8b_2\pi(m+1)}{45h_0^5(m+3)^{3/2}(4m-3)}(\frac{5}{2})B^{\frac{3}{2}}+C_1(\frac{m+3}{2m})B^{\frac{m+3}{2m}-1})\\\nonumber
&-&\frac{1}{2}(-6h_0^2e^{2mN}-\frac{(m-3)B\log(B)(3h_0^2-8\pi
b_1)}{3h_0^2(m-3)^2}-\frac{16b_1m\pi-3h_0^2(C_2m^2-6C_2m+9C_2+4m)B}{3h_0^2(m-3)^2}\\\label{p9*}
&+&\frac{16\sqrt{0.67}8b_2\pi(m+1)}{45h_0^5(m+3)^{3/2}(4m-3)}B^{\frac{5}{2}}+C_1B^{\frac{m+3}{2m}}+16b_0\pi)\}]^{-1}.
\end{eqnarray}
The graphical illustration of energy constraints behavior and
effective EoS parameter is presented in Figure \textbf{8}.
\begin{figure}
\center\epsfig{file=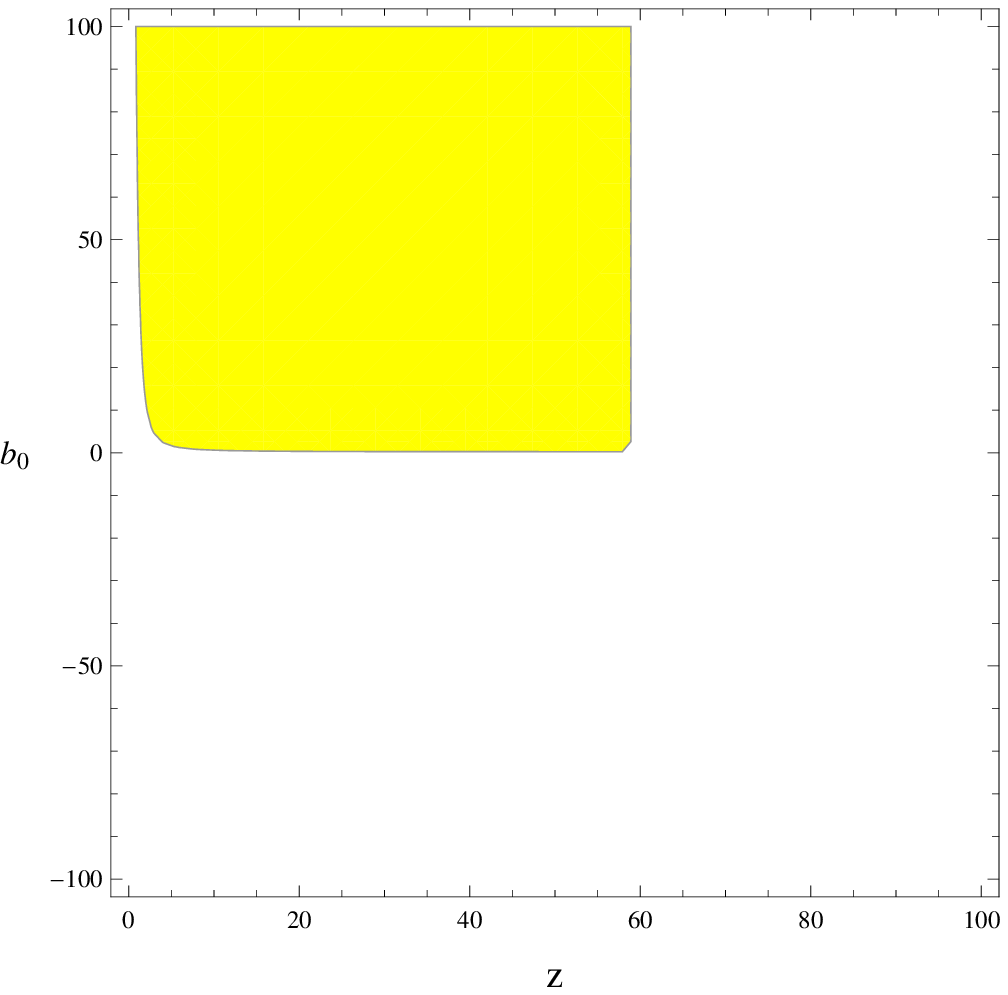, width=0.35\linewidth}
\epsfig{file=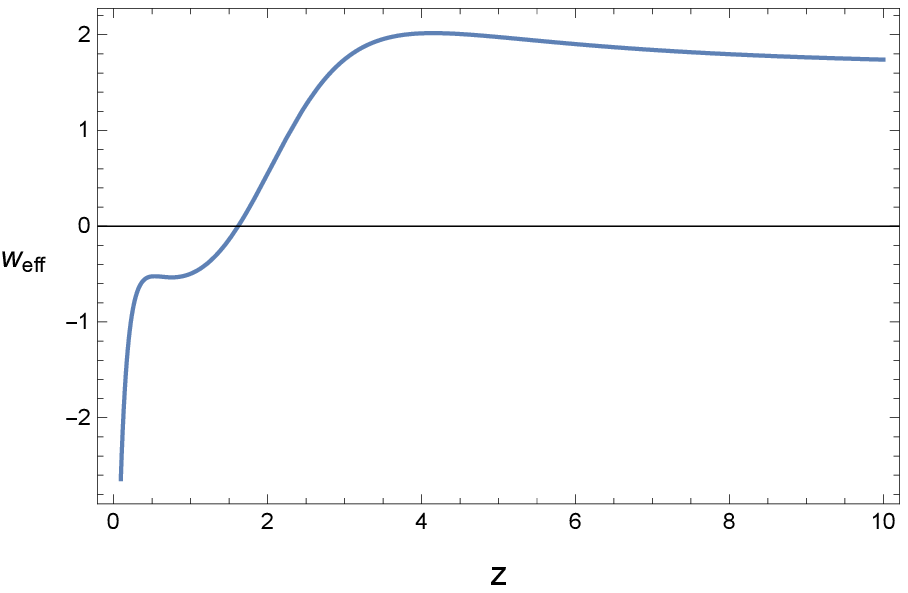, width=0.35\linewidth}\caption{Left region
plot shows validity region for WEC and NEC while Right plot
indicates the evolution of effective EoS parameter versus cosmic
time. For left plot, we take
$b_1=2,~b_2=-0.1,~C_2=C_1=C_3=1,~m=2,~h_0=0.7,~q=-1,~j=1.02$ and
$\omega_m=0$.}
\end{figure}
It is clear form the graph that the WEC and NEC are consistent for
small values of redshift function, i.e., $z\leq60$ with only
positive values of $b_0$. The effective EoS parameters graph shows
that......\\

\section{Conclusion}

Recently, teleparallel theory of gravity and its modifications have
attained significant attention of the researchers for discussing
various issues in cosmology. In the present paper, we have discussed
the energy constraints validity and the effective EoS parameter
evolution in a modified teleparallel gravity namely $f(T,B)$ theory.
Actually, the $f(T,B)$ theory is formulated with the aim to unify
both $f(R)$ and $f(T)$ gravitational frameworks and thus to see how
these theories are connected with each other \cite{27}. It is found
that under some specified limits, this theory reduces to $f(T)$ and
$f(R)$ theories. For discussing the possible constraints on the free
parameters, we have used some famous cosmological models obtained by
the reconstruction scheme in a recent paper \cite{28}. Firstly, we
have derived the general energy conditions directly from the
effective energy-momentum tensor under the transformation
$\rho\rightarrow\rho_{eff}$ and $p_{eff}\rightarrow{p}_{eff}$. Then
we explore the particular forms of these constraints using the
reconstructed $f(T,B)$ function for four different cosmological
models namely: De Sitter universe, power law cosmology, $\Lambda$CDM
universe and Phantom universe model. It is seen that in every case,
there are many free variables that need to be fixed. We have chosen
some specific values for some of these free parameters while others
are restricted graphically in order to make sure the consistency of
these energy bounds. Furthermore, we have used another particular
form of the function given by $f(T,B)=-T+F(B)$ and discussed the
validity of WEC and NEC in terms of graphical regions by fixing the
free parameters in all four cases. The obtained results can be
summarized in the form of the following table:\\

\textbf{Table 1:} The ranges of free parameters for the validity of WEC and NEC obtained through graphs.\\
\begin{table}[bht]
\centering
\begin{small}
\begin{tabular}{|c|c|c|c|}
\hline\textbf{Model}&\textbf{Free Parameters
Choice}&\textbf{Obtained Ranges of free parameters for
Validity}\\
\hline\textbf{$De Sitter Universe Case
I$}&\textbf{$K=1,~\omega_m=0,~\rho_0=1$}&$-20\leq f_0\leq 200,~ 90\leq\tilde{f_0}\leq 170$\\
\hline\textbf{$Case
II$}&$H_0=0.718,~\rho_0=0.001,~\omega_m=0$&$C_1<0,~-1\leq q\leq 2$\\
\hline\textbf{$Power law Model Case
I$}&$\omega_m=0,~t_0=\rho_0=C_2=C_3=1,$&$0\leq C_1\leq200,~0\leq t\leq40$\\
&$h=10$&\\
\hline\textbf{$Case II$}&$\rho_0=t_0=C_2=C_3=1,$& Small $t$ values
with $-100\leq C_1\leq100$\\
&$\omega_m=0,~h=10$& Also, $C_1\leq0$ and $30\leq t\leq50$\\
\hline\textbf{$\Lambda CDM Case I$}&$l=\frac{8\pi}{3},~C_2=10,~\omega_m=0,~a_0=1=\rho_0,$&$C_1>0$ and $z>0$\\
&$C_2=C_3=10$&\\
\hline\textbf{$Case II$}&$\rho_0=t_0=C_2=1,~C_3=-0.1,$&$z>0,~C_1<0$\\
&$\omega_m=0,~l=\frac{8\pi}{3},~C_2=10,~q=-0.64$&\\
\hline\textbf{$Phantom Model Case I$}&$h=0.1,~\omega_m=0,~m=1,~j=1.02,~q=-1,$&$b_0>0,~z\geq3$\\
&$b_1=b_2=-50,~C_1=C_2=C_3=1$&\\
\hline\textbf{$Case
II$}&$b_1=b_2=-50,~C_2=10,~C_1=C_3=1,$&$z\leq60,~b_0>0$\\
&$m=1,~h_0=0.1,~q=-1,~j=1.02,~\omega_m=0$&\\
\hline
\end{tabular}
\end{small}
\end{table}\\

Further we have discussed the evolution of effective EoS parameter
versus cosmic time $t$ or redshift function $z$ by fixing the free
parameters in all four cases graphically. The choices of the free
parameters used here are exactly same as either we used in validity
region graphs of WEC and NEC or obtained through the graphs. The
obtained ranges of EoS parameter are then compared with the
observational ranges as discussed in literature and summarized in
the form of the table \textbf{2}. It is seen that in all cases, the
universe model either corresponds to quintessence era or phantom
cosmic era of universe evolution and $\Lambda$CDM phase. Our results are consistent with
WMAP9 observational data \cite{hinshaw}; $1.073^{+0.090}_{-0.089}$ (WMAP+eCMB+BAO+H$_0$), $1.084\pm0.0063$ (WMAP+eCMB+BAO+H$_0$)
and latest Planck results \cite{ade} $\omega=1.09\pm0.17$ (95\% Planck+WP+Union 2.1).

\newpage

\textbf{Table 2:} Comparison of obtained ranges of effective EoS
parameter
with the observed ranges given in literature.\\
\begin{table}[bht]
\centering
\begin{small}
\begin{tabular}{|c|c|c|c}
\hline\textbf{Model}&\textbf{Free Parameters
Choice}&\textbf{Effective EoS Range}\\
\hline\textbf{$De Sitter Universe Case
I$}&\textbf{$K=1,~\omega_m=0,~\rho_0=1,~\tilde{f}_0=-2$}&$-1.06\leq\omega_{eff}\leq-1$\\
\hline\textbf{$Case II$}&$H_0=0.718,~\rho_0=0.001,~\omega_m=0$&$-1.0163\leq\omega_{eff}\leq-1.0159$\\
\hline\textbf{$Power law Model Case
I$}&$\omega_m=0,~t_0=\rho_0=C_2=C_3=1,$&$-0.968\leq\omega_{eff}\leq-0.962$&\\
&$h=10,~C_1=100$&&\\
\hline\textbf{$Case II$}&$\rho_0=t_0=C_2=C_3=1,$&$-1.05\leq\omega_{eff}\leq-1.25$&\\
&$\omega_m=0,~h=10,~C_1=10$&\\
\hline\textbf{$\Lambda CDM Case I$}&$l=\frac{8\pi}{3},~C_2=10,~\omega_m=0,~a_0=1=\rho_0,$&$-0.998622\leq\omega_{eff}\leq-0.998610$&\\
&$C_1=1000,~C_2=C_3=10$&&\\
\hline\textbf{$Case II$}&$\rho_0=t_0=C_2=1,~C_1=100,~C_3=-0.1,$&$-1.00674\leq\omega_{eff}\leq-1.00666$&\\
&$\omega_m=0,~l=\frac{8\pi}{3},~C_2=10,~q=-0.64$&&\\
\hline\textbf{$Phantom Model Case I$}&$h_0=-0.1,~\omega_m=0,~m=2,~j=1.02,~q=-1$
&$\omega_{eff}<-1$&\\
&$b_0=1,~b_1=2,~b_2=0.0001,~C_1=C_2=C_3=1$&\\
\hline\textbf{$Case II$}&$h_0=0.7,~\omega_m=0,~m=2,~j=1.02,~q=-1$&$\omega_{eff}<-1$&\\
&$b_0=1,~b_1=2,~b_2=-0.1,~C_1=C_2=C_3=1$&\\
\hline
\end{tabular}
\end{small}
\end{table}\\
It would be interesting to investigate the possible constraints on
the free parameters using matter density perturbations for some other
well-known models of cosmology in $f(T,B)$ gravity.

\vspace{.5cm}

\section*{Acknowledgments}

``M. Zubair thanks the Higher Education Commission, Islamabad, Pakistan for its
financial support under the NRPU project with grant number
$\text{7851/Balochistan/NRPU/R\&D/HEC/2017}$''.

\end{document}